\documentclass[aps,prb,amsmath,amssymb,footinbib,showpacs,twocolumn,superscriptaddress]{revtex4-2}
\usepackage{amsmath}
\usepackage{amssymb}
\usepackage{amsthm}
\usepackage{setspace}
\usepackage{graphicx}
\usepackage{braket}
\usepackage{mathrsfs}
\usepackage{float}
\usepackage[colorlinks = true,linkcolor = red,urlcolor  = blue,citecolor = blue,anchorcolor = blue]{hyperref}
\usepackage[utf8]{inputenc}
\usepackage[english]{babel}
\usepackage{bm}

\begin{document}

\title{Proximity-induced superconductivity generated by thin films: Effects of Fermi surface mismatch and disorder in the superconductor}

\author{Tudor D. Stanescu}
\affiliation{Department of Physics and Astronomy, West Virginia University, Morgantown, WV 26506, USA}
\author{Sankar Das Sarma}
\affiliation{Condensed Matter Theory Center and Joint Quantum Institute, Department of Physics, University of Maryland, College Park, MD, 20742-4111, USA}

\begin{abstract}
We investigate the effects of disorder characterising a superconducting thin film on the proximity-induced superconductivity generated by the film (in, e.g., a semiconductor) based on the exact numerical analysis of a three-dimensional microscopic model. To make the problem numerically tractable, we use a recursive Green's function method in combination with a ``patching approach'' that exploits the short-range nature of the interface Green's function in the presence of disorder. As a result of the Fermi surface mismatch between the superconductor (SC) and the semiconductor (SM) in combination with the confinement-induced quantization of the transverse SC modes, the proximity effect induced by a clean SC film is typically one to three orders of magnitude smaller that the corresponding quantity for a bulk SC and exhibits huge thickness-dependent variations. The presence of disorder has competing effects: on the one hand it enhances the proximity-induced superconductivity and suppresses its strong thickness dependence, on the other hand it generates proximity-induced effective disorder in the SM. The effect of proximity-induced disorder on the topological superconducting phase and the associated Majorana modes is studied nonperturbatively. 
\end{abstract}

\maketitle

\section{Introduction}\label{S1}

The superconducting proximity effect, where a parent superconductor (e.g. Al, Nb, Pb) induces superconductivity into a nearby non-superconductor (a non-superconducting metal or a semiconductor), is a fascinating physical phenomenon arising from the tunneling of Cooper pairs from the superconductor into the proximate non-superconductor through a contact interface.  Proximity-induced superconductivity has been known for 90 years \cite{Holm1932} and has been well-studied experimentally \cite{Meissner1960,vanSon1987}.
A hybrid system of particular  interest is a metallic superconductor (often Al) in contact with a semiconductor, which generates proximity-induced superconductivity in the semiconductor \cite{vanHuffelen1993,JarilloHerrero2006}. The ``textbook'' proximity effect \cite{deGennes2018} involving a semi-infinite superconductor in contact with a semi-infinite non-superconducting metal is reasonably well understood \cite{deGennes1964}. It basically describes a process where the non-superconductor develops a finite-size region with proximity-induced superconductivity near the interface, over a distance of the order of the coherence length, through Cooper pair tunneling across the boundary. In this case, there is no superconductivity induced in the bulk of the non-superconductor.  This boundary superconductivity proximity-induced in bulk (non-superconducting) systems by a bulk superconductor is significantly different from the subject matter of our work, which concerns the proximity effect generated by a thin film parent superconductor proximity-coupled to a finite size quasi one dimensional (1D) or quasi two dimensional (2D) non-superconductor (in our case a semiconductor). In this case, both components of the hybrid structure are small finite size subsystems and, typically, the whole non-superconductor acquires proximity-induced superconductivity. This proximity phenomenon is often thought of in terms of Andreev processes in a mesoscopic superconductor \cite{Andreev1964,Blonder1982,Beenakker2000,Stanescu2010}.
  
Interest in the proximity effect generated by thin films in superconductor (SC)-semiconductor (SM) hybrid  systems has increased enormously over the past decade as a result of concrete theoretical proposals predicting the possibility of inducing topological superconductivity and realizing localized Majorana bound states (and the associated Majorana zero modes) in a SM with strong spin-orbit coupling proximity-coupled to a conventional (non-topological) SC \cite{Sau2010,Sau2010a,Lutchyn2010,Oreg2010}.
 The subject is vast, with thousands of experimental and theoretical publications since 2010. Here, we focus on a crisp theoretical understanding of the role of disorder (inside the SC) in proximity-inducing superconductivity using thin SC films. Hence, we do not provide a review of the vast subject dedicated to the search for Majorana zero modes (MZMs) in SC-SM structures, its successes and failures, and the world-wide current activity, which has been partially reviewed elsewhere \cite{Alicea2012,Leijnse2012,Stanescu2013,Beenakker2013,DSarma2015,Sato2017,Lutchyn2018,Prada2020}.
 Instead, we concentrate on the specific role played by the disorder present inside a thin SC film (particularly the surface disorder) in the superconducting proximity effect induced by the film, a subject that has attracted  little attention in the literature in spite of its significant potential importance.

After the great initial excitement generated by the experimental observation \cite{Mourik2012,Deng2012,Das2012,Churchil2013,Finck2013}
of zero-bias conductance peaks, which are predicted as possible signatures of Majorana bound states in SC-SM structures, it was realized that disorder is playing an extremely important role in the system \cite{Kells2012,Akhmerov2011,Liu2012,Bagrets2012,Sau2013,Adagideli2014,Moore2018,Pan2020a} and, most likely, is essentially preventing the predicted emergence of topological superconductivity in experimentally available structures. 
In particular, in all early experiments the proximity-induced superconducting gap was extremely soft, i.e. there was substantial weight associated with subgap fermionic states. This was explained as arising from disorder being present at the SM-SC interface \cite{Takei2013} and, according to this interpretation, the soft gap feature should disappear (leading to a hard gap) if the SC-SM interface is clean, e.g., if it is an epitaxial interface with no disorder. Subsequent efforts resulted in the growth of epitaxial SC-SM interfaces, which, indeed, led to a hard proximity-induced gap \cite{Chang2015}.  More recently, however, it has been realized through extensive simulations that disorder inside the semiconductor (due to, e.g., the presence of charge impurities) is generating low energy 
Andreev bound states (ABSs) in the system \cite{Pan2020,DasSarma2021,Ahn2021,Woods2021,Zhang2021}, which are most likely responsible for the zero bias conductance peaks observed experimentally.
 Thus, disorder has emerged as the main enemy of topological proximity-induced superconductivity and needs to be understood thoroughly.

In contrast to the SM disorder, which has been extensively discussed in the literature, very little has been done on understanding the role of disorder that may be present inside the parent superconducting thin film.  This is a rather surprising omission since the disorder in the metallic SC film (made of Al in most of the current SC-SM structures) would be much stronger than the SM disorder because the SC film is highly amorphous, with an electron mean free path that is likely to be limited by the film thickness ( $<10~$nm), whereas the corresponding SM mean free path is about an order of magnitude larger. Nonetheless, the (relatively weaker) SM disorder is now known to be extremely detrimental to the realization of topological superconductivity.  Hence, why this neglect of understanding the effects of SC disorder on the superconducting proximity effect, even if the SC disorder is extremely strong (i.e. much stronger than any SM disorder)?  
In fact, an early work \cite{Potter2011} did claim that ``the induced superconductivity is strongly susceptible to disorder'' in the SC, but it turned out that this claim was technically incorrect \cite{Potter2011a}. On the other hand, a correct perturbative calculation established the immunity of the SM proximity effect to (weak) SC disorder \cite{Lutchyn2012}. However, the actual SC disorder, which is characterized by an energy scale $> 10-100~$meV, is orders of magnitude larger than the proximity-induced SM gap ($\sim 0.1~$ meV), so the diagrammatic weak-coupling arguments of Ref. \onlinecite{Lutchyn2012} may not apply. Considering the  demonstrated importance of the (relatively weaker) SM disorder and the nonperturbative nature of the SC disorder, it is crucial to understand in detail the role of SC disorder, both qualitatively and quantitatively.  

There is, however, another important problem regarding the proximity effect generated in SC-SM hybrid structures by thin SC films: the ``Fermi surface mismatch'' between the SC, which is characterized by a Fermi energy  $\sim 10~$ eV, and the SM, which has a Fermi energy on the order $\sim 10-100~$meV. The relatively small SM Fermi energy corresponds to long wavelength SM states that, in a clean hybrid system, will only couple to SC states characterized by small values of the wave vector ${\bm k}_\parallel$ parallel to the SM-SC interface. To generate proximity-induced superconductivity, these long wavelength SC states should also have low energy (comparable to the SC gap).
However, due to the finite size quantization of the transverse modes in thin SC films such states may not be available, except for fine tuned values of the film thickness. In other words, in a clean thin SC film - SM structure, the Fermi surface mismatch combined with the strong finite size quantization of the transverse modes makes it typically impossible for SC proximity effect to manifest itself, because electrons cannot tunnel coherently through the SC-SM interface. Furthermore, the induced proximity effect has an extremely strong dependence of the film thickness, with orders of magnitude variations corresponding to changes in the film thickness by one atomic layer. To the best of our knowledge, the effect of SC disorder on this phenomenon has not been properly studied in the literature. 

In this work, we address both issues regarding the Fermi surface mismatch in thin SC films and  nonperturbative disorder effects by exactly solving the SC problem at the mean field level, starting with a lattice Bogoliubov-de Gennes (BdG) Hamiltonian that explicitly incorporates disorder. By focusing on disorder located near the free surface of the thin SC film, i.e. away from the SC-SM interface, we show that SC disorder has competing effects: on the one hand, the presence of SC disorder helps generate robust induced superconductivity by suppressing (or even eliminating) the Fermi surface mismatch problem; on the other hand, SC disorder induces effective disorder in the (active) SM component, which can be detrimental to the realization of topological superconductivity. 
We demonstrate that, remarkably, (i) there exist surface disorder models consistent with the realization of robust induced superconductivity -- i.e., there are  disordered thin SC films that can generate a proximity effect that is comparable with the effect generated by a bulk superconductor and has a weak dependence on the film thickness -- and (ii) the corresponding induced effective disorder is consistent with the presence of topological superconductivity, provided that the SC-SM coupling is weak enough. This is an important result that, on the one hand, demonstrates the feasibility of robust topological superconductivity in SM-SC systems using thin SC films, and, on the other hand, suggests that a detailed experimental study of the SC properties is critical to optimizing the structure and enhancing the stability of the topological phase.   

The remainder of this paper is organized as follows. In Sec. \ref{S2} we present our theoretical model and the recursive method for efficiently calculating the Green's function of the superconducting thin film at the interface. The numerical results are discussed in  Sec. \ref{S3}, starting with the clean case in Sec. \ref{S3A}, then considering a superconducting thin film with surface disorder in  Sec. \ref{S3B}. We specifically consider two models of surface disorder, one consisting of a random onsite potential within a thin layer ($\sim 2~$nm) near the surface (Sec. \ref{S3B1}), the other representing a model of surface roughness (Sec. \ref{S3B2}). Next, in Sec. \ref{S4}, we assume that the superconducting film with surface roughness is proximity-coupled to a semiconductor nanowire and we determine the effect of the (SC) surface disorder on the low-energy physics of the wire. Our conclusions are discussed in Sec. \ref{S5}. Details regarding a few important technical aspects and additional numerical results are provided in Appendixes \ref{AppA}-\ref{AppD}.

\section{Theoretical model}\label{S2}

\begin{figure}[t]
\begin{center}
\includegraphics[width=0.48\textwidth]{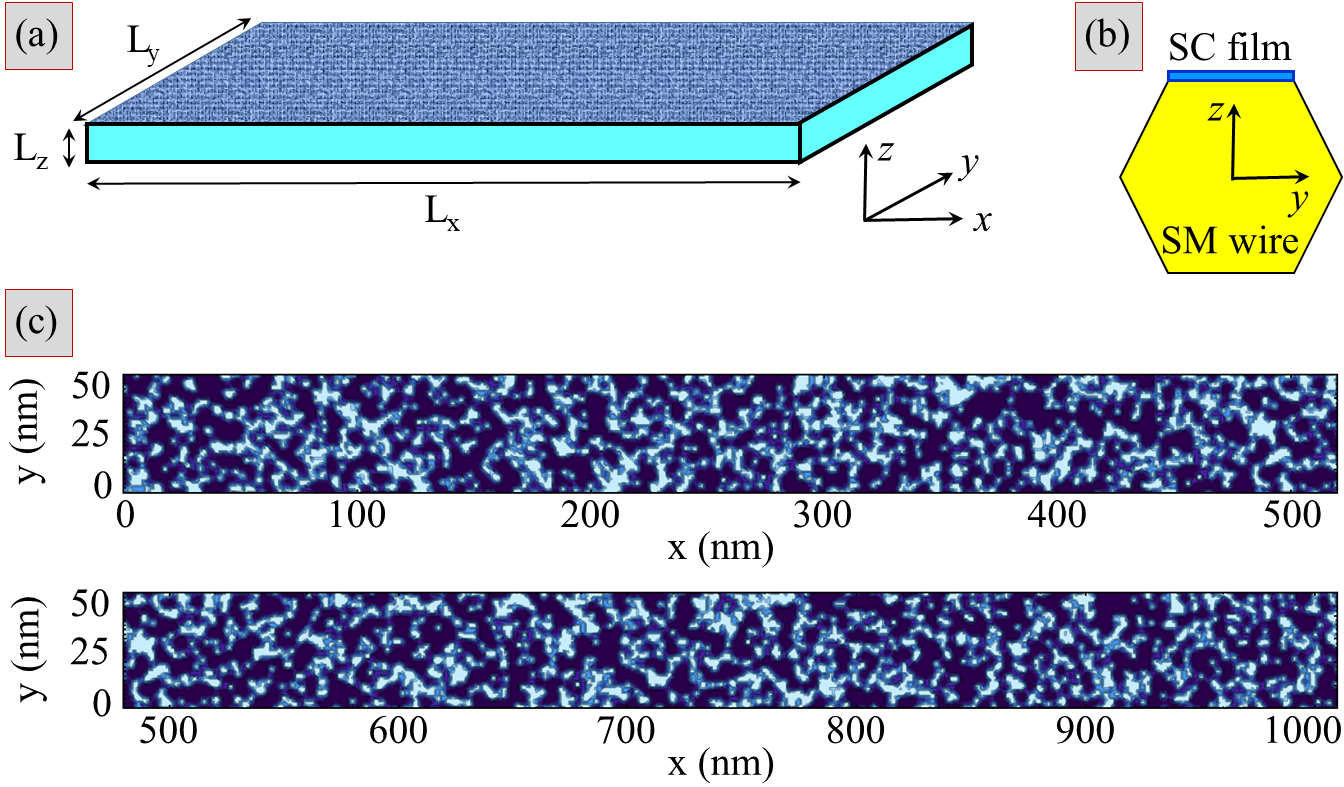}
\end{center}
\vspace{-3mm}
\caption{(a) Schematic representation of a finite 3D superconducting film with surface roughness on the top surface. (b) Example of proximity-coupled semiconductor-superconductor (SM-SC) structure containing a superconducting thin film. (c) Specific surface roughness realization used in the calculations  (see Sec. \ref{S3B2}  and Sec. \ref{S4}). The parameter values are: $L_x= 1.025~\mu$m, $L_y =  51.3~$nm, and $L_z= 6.25 \pm 0.35~$nm. Different shades of blue correspond to local variations of the film thickness affecting the top three atomic layers. Note that  the wire is broken into two segments, for clarity.}
\label{FIG1}
\vspace{-1mm}
\end{figure}

Our main goal is to understand the effect of disorder characterizing a thin superconducting film on the superconducting proximity effect induced by the film [see, e.g., Fig. \ref{FIG1}]. Within a Green's function approach, after ``integrating out'' the degrees of freedom of the SC subsystem, the  proximity effect is described by a self-energy contribution to the Green's function of the ``active component'', e.g., a SM nanowire [see Fig. \ref{FIG1}(b)] or a quasi-2D electron gas hosted by a SM quantum well \cite{Shabani2016}. In turn, this self-energy is proportional to the Green's function $G_{SC}$ of the superconductor at the interface with the ``active component'' (e.g., the SC-SM interface), 
\begin{equation}
{G}_{SC}(\omega; \widetilde{\bm r}, \widetilde{\bm r}^\prime) \equiv \widetilde{G}(\omega; {\bm r}_\parallel, {\bm r}_\parallel^\prime) =\left[ (\omega - H_{SC})^{-1}\right]_{\widetilde{\bm r}, \widetilde{\bm r}^\prime},   \label{Eq1}
\end{equation}
where $\widetilde{\bm r} = (x, y, \widetilde{z})$ is the 3D position vector of a point at the interface defined by the condition $z=\widetilde{z}$ and  ${\bm r}_\parallel = (x, y)$ is the corresponding 2D position vector. In Eq. (\ref{Eq1}) the superconductor is described at the mean-field level by the Bogoliubov-de Gennes Hamiltonian  $H_{SC}$.  More specifically, we use a tight-binding model defined on a simple tetragonal lattice with lattice parameters $a$ (in the $x-y$ plane) and $c$ (along the $z$ direction) and having the second quantized form
\begin{eqnarray}
H_{SC} &=& -\sum_{\langle{\bm i}, {\bm j}\rangle} \sum_{\sigma} t_{{\bm i} {\bm j}}~\! \hat{c}_{{\bm i}\sigma}^\dagger \hat{c}_{{\bm j}\sigma}^{~} +\sum_{{\bm i}, \sigma} [V({\bm r}_{\bm i}) -\epsilon_F]\hat{c}_{{\bm i}\sigma}^\dagger \hat{c}_{{\bm i}\sigma}^{~} \nonumber \\
&+& \sum_{{\bm i}}\Delta_0\left(\hat{c}_{{\bm i}\uparrow}^\dagger \hat{c}_{{\bm i}\downarrow}^\dagger +   \hat{c}_{{\bm i}\downarrow}^{~}\hat{c}_{{\bm i}\uparrow}^{~}\right), \label{Eq2}
\end{eqnarray}
where ${\bm i} = (i_x, i_y, i_z)$ labels the lattice site with position vector ${\bm r}_{\bm i} = (a~\! i_x, a~\! i_y, c~\! i_z)$, $\hat{c}_{{\bm i}\sigma}^\dagger$ ($\hat{c}_{{\bm i}\sigma}^{~}$) is the creation (annihilation) operator for an electron with spin $\sigma$ located at site ${\bm i}$, 
$\langle{\bm i}, {\bm j}\rangle$ are nearest-neighbor sites satisfying the condition $|{\bm i}- {\bm j}|=1$, and $t_{{\bm i} {\bm j}}$ are the corresponding hopping matrix elements, with $t_{{\bm i} {\bm j}} = t$ if $|{\bm r}_{\bm i} - {\bm r}_{\bm j}| = a$ (hopping in the $x-y$ plane) and $t_{{\bm i} {\bm j}} = t_z$ if $|{\bm r}_{\bm i} - {\bm r}_{\bm j}| = c$ (hopping along the $z$ direction). The position-dependent quantity $V({\bm r}_{\bm i})\equiv V_{\bm i}$  represents a disorder-generated potential, $\epsilon_F$ is the Fermi energy of the SC, and $\Delta_0$ is the pairing potential. Note that we neglect the possible position dependence of $\Delta_0$ in the presence of disorder and we do not consider the effect on $\Delta_0$ of an applied magnetic field. These effects can be incorporated into the theory by solving self-consistently a position-dependent gap equation, which involves a substantial numerical cost and is beyond the scope of this work. 
The key physical phenomenon of interest in this work, which involves the role of SC disorder in the proximity effect induced by a thin SC film, is not affected in any qualitative manner by these additional complications.

 Next, we note that, due to the absence of spin-orbit coupling, the SC  Hamiltonian matrix has the block form
 \begin{equation}
 H_{SC} = \left(
 \begin{array}{cccc}
 H_n & 0 & 0 & \Delta_0 I \\
 0 & H_n & - \Delta_0 I & 0 \\
 0 & -\Delta_0 I & - H_n & 0 \\
 \Delta_0 I & 0 & 0 & -H_n
 \end{array}\right), \label{Eq3}
 \end{equation}  
 with $H_n$ being the Hamiltonian matrix that describes the spin-degenerate electrons in the normal phase, 
 \begin{equation}
 [H_n]_{{\bm i} {\bm j}} = -t_{{\bm i} {\bm j}} + (V_{\bm i}-\epsilon_F)\delta_{{\bm i} {\bm j}},   \label {Eq4}
\end{equation}
while $0$ and $I$ are the zero and identity square matrices, respectively, of dimension $N=N_x N_y N_z$ given by the number of sites in the 3D lattice.
This simple spin structure is inherited by the interface Green's function in Eq. (\ref{Eq1}), hence it is enough to focus on one nontrivial block (e.g., the one containing the $+\Delta_0 I$ contributions). Nonetheless, inverting the corresponding $2N\times 2N$ matrix can be numerically challenging if the number of degrees of freedom is large (e.g., millions of sites). To make the numerical procedure more efficient, we use a recursive Green’s function method \cite{Thouless1981,MacKinnon1985} 
  that involves solving the set of equations
\begin{eqnarray}
G^{(\ell)}(\omega) &=& \left[
\left(\begin{array}{cc}
\omega I- H_n^{(\ell)} & -\Delta_0 I \\
 -\Delta_0 I &  \omega I+ H_n^{(\ell)}
\end{array}\right) - \Sigma^{(\ell)}(\omega))\right]^{-1}       \label{Eq5} \\
 \Sigma^{(\ell)}(\omega) &=& 
 \left(\begin{array}{cc}
-t_z I & 0 \\
0 & t_z I
\end{array}\right) G^{(\ell-1)}(\omega)
\left(\begin{array}{cc}
-t_z I & 0 \\
0 & t_z I
\end{array}\right),            \label{Eq5}
\end{eqnarray}
where $H_n^{(\ell)}$ is the Hamiltonian matrix corresponding to  layer $\ell$, i.e., having elements $ [H_n]_{{\bm i} {\bm j}}$ with $i_z = j_z = \ell$, and $I$ is the $N_\parallel\times N_\parallel$ identity matrix, with $N_\parallel= N_x N_y$. The recursive procedure is repeated $N_z$ times starting with $\Sigma^{(1)}$ being equal to the $2N_\parallel\times2N_\parallel$ zero matrix. Finally, we have $\widetilde{G}(\omega) = G^{(N_z)}(\omega)$.

Before presenting our main results, we point out that the problem of calculating the interface Green's function simplifies considerably in the case of a clean system ($V_{\bm i} = 0$). Indeed, after performing a 3D discrete Fourier transform, we obtain 
\begin{eqnarray}
\widetilde{G}(\omega, \epsilon_{n_x n_y}) &=& - \gamma(\omega,\epsilon_{n_x n_y})
\left(\begin{array}{cc}
\omega & -\Delta_0  \\
 -\Delta_0  &  \omega
\end{array}\right)  \nonumber \\
&+& \zeta(\omega,\epsilon_{n_x n_y})
\left(\begin{array}{cc}
1 & 0  \\
0 & -1
\end{array}\right),      \label{Eq7}
\end{eqnarray}
where
\begin{eqnarray}
\gamma(\omega,\epsilon_{n_x n_y}\!) &=& \frac{2}{N_z\!+\!1}\! \sum_{n_z=1}^{N_z}\!\frac{\sin^2\!\left(\frac{n_z \pi}{N_z + 1}\right)}{E_{n_x n_y n_z}^2 \!+\!\Delta_0^2\!-\!\omega^2}, \label{Eq8} \\
\zeta(\omega,\epsilon_{n_x n_y}\!) &=& \frac{2}{N_z\!+\!1}\! \sum_{n_z=1}^{N_z}\!\frac{E_{n_x n_y n_z}~\!\sin^2\!\left(\frac{n_z \pi}{N_z + 1}\right)}{E_{n_x n_y n_z}^2 \!+\!\Delta_0^2\!-\!\omega^2}. \label{Eq9}
\end{eqnarray}
The energy of the mode $(n_x n_y n_z)$ is $E_{n_x n_y n_z} = \epsilon_{n_z} +\epsilon_{n_x n_y} -\epsilon_F$, where 
\begin{eqnarray}
\epsilon_{n_z} &=& 2t_z\left(1-\cos\frac{n_z \pi}{N_z+1}\right), \label{Eq10} \\
\epsilon_{n_x n_y} &=& 2t\left(2-\cos\frac{n_x \pi}{N_x+1}-\cos\frac{n_y \pi}{N_y+1}\right). \label{Eq11}
\end{eqnarray}
The quantity $\gamma(\omega,\epsilon_{n_x n_y}\!)$ is related to the (normal phase)  interface density of states of the mode $(n_x n_y)$ at the Fermi energy, 
\begin{equation}
\widetilde{\nu}_{_F}(\epsilon_{n_x n_y}) = -\frac{1}{\pi}{\rm Im}\left[ \frac{2}{N_z\!+\!1}\!\sum_{n_z=1}^{N_z} \frac{\sin^2\!\left(\frac{n_z \pi}{N_z + 1}\right)}{\omega-E_{n_x n_y n_z}+i\eta}\right]_{\omega=0}.               \label{Eq12}
\end{equation}
More specifically, we have 
\begin{equation}
\gamma(\omega, \epsilon) = \frac{\pi}{\sqrt{\Delta_0^2 - \omega^2}}\widetilde{\nu}_{_F}(\epsilon)\vert_{\eta=\sqrt{\Delta_0^2-\omega^2}}. 
\end{equation}
For an infinitely-thick superconductor, $N_z \rightarrow \infty$, the summation in Eq. (\ref{Eq12}) cand be done explicitly and we have 
\begin{equation}
\widetilde{\nu}_{_F}^\infty(\epsilon_{n_x n_y}) = \frac{1}{\pi t_z}\sqrt{1-\frac{(2t_z +\epsilon_{n_x n_y} -\epsilon_F)^2}{4 t_z^2}}. \label{Eq14}
\end{equation}
At this point, it is important to remember that our goal is to understand the proximity effect induced by the superconductor when coupled to another subsystem (e.g., a semiconductor) across the interface. Focusing on Majorana SM-SC hybrid structures, we note that the relevant SM energy scales are on the order of meV, up to tens of meV if we assume high subband occupancy \cite{Woods2020}, which implies characteristic wave vectors smaller than about $0.25~$nm$^{-1}$ for an InAs-based SM-SC structure. Consequently, the relevant in-plane SC modes, which are characterized by comparable values of the (in-plane) wave vector,  ${\bm k}_\parallel(n_x, n_y) = (\pi n_x/L_x, \pi n_y/L_y)$, have a maximum energy scale on the order of $1-2~$meV. On the other hand, $t_z$ and $\epsilon_F$ have typical values on the order of $10~$eV. We conclude that the dependence on  $\epsilon_{n_x n_y}$ in Eq. (\ref{Eq14}) is negligible, so that the interface density of states for an infinitely-thick SC slab is practically described by a constant, $\widetilde{\nu}_{_F}^\infty \equiv \widetilde{\nu}_{_F}^\infty(0)$. We choose this constant as the ``natural unit'' for the interface Green's function and we redefine the ``amplitudes'' $\gamma$ and $\zeta$ in Eq. (\ref{Eq7}) in terms of the dimensionless quantities $\tilde{g}$ and $\tilde{v}$ as
\begin{equation}
\gamma(\omega,\epsilon) = \frac{\pi \nu_{_F}^\infty}{\sqrt{\Delta_0^2-\omega^2}}~\!\widetilde{g}(\omega,\epsilon), ~~~~~~\zeta(\omega,\epsilon) = \pi \nu_{_F}^\infty~\!\widetilde{v}(\omega,\epsilon). \label{Eq15}
\end{equation}
The key quantity $\widetilde{g}(\omega,\epsilon)$ provides a measure of the ``strength'' of the superconducting proximity effect generated by the SC film relative to the effect induced by an infinitely thick superconductor. Of course, we have $\widetilde{g}(\omega,\epsilon) \xrightarrow[~]{N_z\rightarrow\infty} 1$. Finally, we note that the block structure of the interface Green's function given in Eq. (\ref{Eq7}) holds even in the presence of disorder, but the quantities $\gamma$ and $\zeta$ (or $\widetilde{g}$ and $\widetilde{v}$) become $N_\parallel \times N_\parallel$ matrices. Hence, in general we have 
\begin{equation}
\widetilde{G}(\omega; {\bm i}, {\bm j}) = \pi \nu_{_F}^\infty\!\! \left[\widetilde{g}_{{\bm i}{\bm j}}(\omega)\frac{-\omega\tau_0 \!+\!\Delta_0\tau_x}{\sqrt{\Delta_0^2-\omega^2}} + \widetilde{v}_{{\bm i}{\bm j}}(\omega) \tau_z\right],  \label{Eq16}
\end{equation}  
with ${\bm i}$ and ${\bm j}$ labeling lattice sites at the interface and $\tau_\mu$ being Pauli matrices associated with the particle-hole degree of freedom.  

The bulk of the numerical results presented below are obtained using the following values of the model parameters: lattice constants $a = 1.22~$nm (in-plane) and $c=1.17~$\AA$~$ (in the $z$ direction), hopping parameters $t=\hbar^2/2m_e a^2 = 25.57~$meV (in-plane) and $t_z= 6.402~$eV (in the $z$ direction), and pairing potential $\Delta_0 = 0.33~$meV. For calculations involving different parameter values we explicitly provide those values.
We note that the chosen values of the in-plane parameters take into account the fact that only the low-lying modes with $k_\parallel \lesssim 0.25~$nm$^{-1}$ are physically relevant. On the other hand, $c$ was chosen as approximately half of the inter-layer spacing for Al along the $(111)$ direction, while $t_z$ was determined by the values of the Fermi k-vector and Fermi velocity for Al,  $k_F= 17.5~$nm$^{-1}$ and $v_F=2.02\times 10^6~$m/s. Note that this gives a Fermi energy $\epsilon_F = 18.7~$eV, larger than the actual value for Al. A more detailed discussion of this parameter choice is provided in the next section. 

\section{Numerical results: The interface Green's function of a thin SC film}\label{S3}

In this section we calculate numerically the interface Green's function $\widetilde{G}$ of a thin SC film based on the theoretical model described above. We focus on the ``relative amplitude'' $\widetilde{g}$, which plays the key role in the superconducting proximity effect, determining the induced pairing potential and the proximity-induced energy renormalization (also see Sec. \ref{S4}), and investigate its dependence on the relevant quantities, including the film thickness. The clean case, which clearly illustrates the major problem associated with the Fermi surface mismatch and the quantization of the transverse modes, is discussed in Sec. \ref{S3A}; the effect of surface disorder is investigated in Sec.\ref{S3B}.

\subsection{The clean case}\label{S3A}

\begin{figure}[t]
\begin{center}
\includegraphics[width=0.48\textwidth]{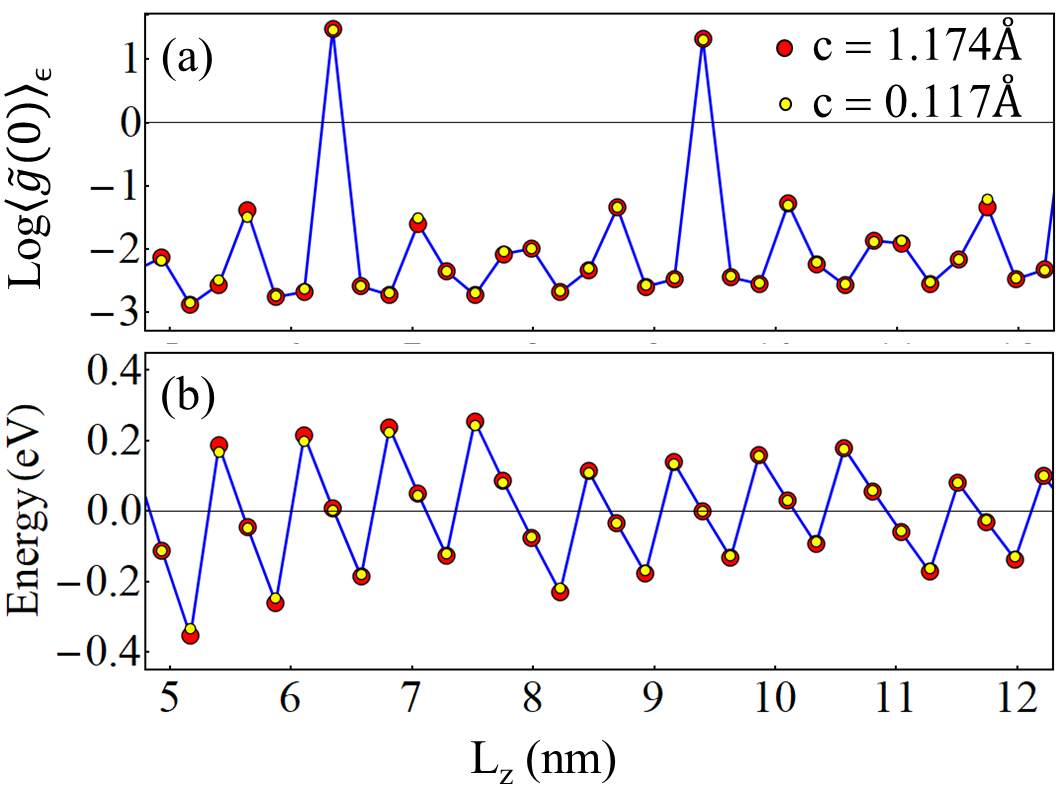}
\end{center}
\vspace{-3mm}
\caption{(a) Dependence of the averaged Green's function ``amplitude'', $\widetilde{g}$, on the SC film thickness, $L_z$, for a clean system. The relative amplitude $\widetilde{g}(0,\epsilon)$ was averaged over an energy window $-10~$meV$~\!<\epsilon<10~$meV. Note that ${\rm Log}[\dots]$ represents the logarithm with base 10. The distance $\delta L_z \approx 2.34~$\AA~ between two successive data points corresponds to one atomic Al $(111)$ layer. Note that SC films with $L_z\approx 6.34~$nm and $L_z\approx 9.4~$nm have an (average) interface Green's function larger that the infinitely thick SC (by about one order of magnitude), while all other thicknesses correspond to values of the interface Green's function 1-3 orders of magnitude smaller that the bulk SC value. (b) Energy of the transverse band minimum, $\epsilon_{n_z}-\epsilon_F$ [Eq. (\ref{Eq10})], closest to the Fermi energy as a function of the film thickness. Note that the maxima in (a) correspond to transverse bands having their minima very close to the Fermi energy. The results for $c=0.117~$\AA~ (yellow dots) were shifted with respect to those corresponding to $c=1.17~$\AA~ (red dots) by $\delta L_z= -0.47~$nm and $\delta \epsilon_{n_z} = 12.5~$meV. }
\label{FIG2}
\vspace{-1mm}
\end{figure}

For a clean SC film, the relative amplitude $\widetilde{g}(\omega, \epsilon)$ of the interface Green's function can be determined using Eqs. (\ref{Eq12}-\ref{Eq15}). Focusing on zero frequency, $\omega =0$, and taking into account the fact that the SC states relevant to the proximity effect have small in-plane energy contributions ($\epsilon_{n_x n_y}\sim 1-5~$meV), we calculate the quantity $\langle \widetilde{g}(0)\rangle_\epsilon$ representing the zero frequency Green's function amplitude averaged over energy within the range $-10~{\rm meV} < \epsilon < 10~{\rm meV}$. The dependence of ${\rm Log}\langle \widetilde{g}(0)\rangle_\epsilon$ on the film thickness $L_z$ is shown in Fig. \ref{FIG2}(a). Note that, except for two specific $L_z$ values, the (average) interface Green's function amplitude is 1-3 orders of magnitude smaller than the corresponding quantity for a bulk SC. Furthermore, $\widetilde{g}$ varies strongly (by up to four orders of magnitude) when the film thickness changes by a single atomic layer. This behavior is due to the absence of low-energy in-plane modes $(n_x, n_y)$ with $|\epsilon|\leq 10~$meV as a result of the large inter-band spacing of the transverse modes $n_z$, $\epsilon_{n_z+1}-\epsilon_{n_z} \sim 200-700~$meV for the $L_z$ range considered here. These low-energy (in-plane) modes have total energies right above the minima of the transverse bands, which are typically far from the Fermi energy, as shown in Fig. \ref{FIG2}(b). Furthermore, the energies of the transverse band minima change (typically by $100-200~$meV) when the film thickness varies by one atomic layer, which explains the strong dependence of $\widetilde{g}$ on the film thickness.
The results in Fig. \ref{FIG2} clearly illustrate the fundamental problem of proximity-inducing superconductivity using thin, clean SC films: confinement-generated quantization (along the $z$ direction)  is inconsistent with the presence of low-lying in-plane modes, except for specific values of the film thickness.  Realizing (clean) SC films that have low-lying modes requires exquisite fine-tuning (and luck, which is less likely to happen in practice).

Before continuing our analysis we address a technical aspect alluded to in the previous section. The hopping in the $z$ direction is determined by the lattice constant, $c$, and the Fermi velocity, $v_F$, of Al (rather than the Fermi energy, $\epsilon_F$). Consequently, for $c=1.17~$\AA~ the resulting Fermi energy is larger that the corresponding Al value. This issue can be addressed  by using a finer grid. For example, with a lattice constant $c=0.117~$\AA~ one can practically match both the Fermi velocity and the Fermi energy of Al and one gets $\epsilon_F \approx  11.7~$eV. Of course, this involves a significant numerical cost. However, the results in Fig. \ref{FIG2} show that paying this numerical cost is not necessary. More specifically, since the two sets of parameters (corresponding to red and yellow dots in Fig. \ref{FIG2}) are characterized by the same $v_F$ value, they generate similar inter-band spacings near the Fermi energy, the only difference being an overall shift by  $\delta L_z= -0.47~$nm and $\delta \epsilon_{n_z} = 12.5~$meV of the yellow points. In other words, as long as we are interested in semi-quantitative results (e.g., that the probability of obtaining relative amplitudes $\widetilde{g}$ of order one or larger within the range of film thicknesses $4 < L_z < 15~$nm is about $7\%$), without  focusing on specific quantitative information (e.g., the exact values of $L_z$ that generate large $\widetilde{g}$), using a larger $c$ value is physically consistent and numerically convenient. 

\begin{figure}[t]
\begin{center}
\includegraphics[width=0.48\textwidth]{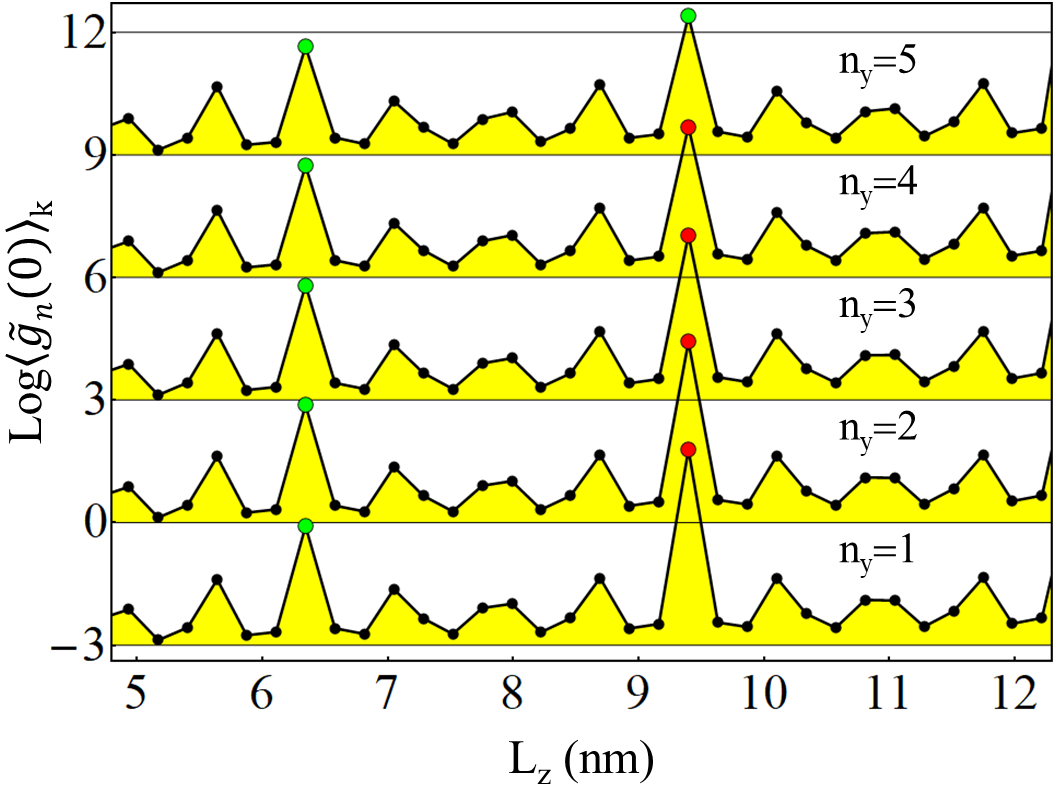}
\end{center}
\vspace{-3mm}
\caption{Dependence of the $k$-averaged $\widetilde{g}_{n_y}(0)$ on the SC film thickness, $L_z$, for a clean system of finite width $L_y=51.3~$nm. The relative amplitude $\widetilde{g}_{n_y}(0, k)$ for the lowest five $n_y$ modes  was averaged over the $k$ range $0 \leq k \leq 0.25~$nm$^{-1}$. Note that the corresponding curves are shifted by $3(n_y-1)$, for clarity. Amplitudes smaller than $1/\sqrt{10}$ (i.e., ${\rm  Log}\langle\widetilde{g}_{n_y}(0)\rangle_k < -0.5$) are marked by black dots, while amplitudes larger than  $\sqrt{10}$ (i.e., ${\rm  Log}\langle\widetilde{g}_{n_y}(0)\rangle_k  > 0.5$) are marked by red dots. The green dots correspond to ``optimal'' amplitudes, $0.316 < \langle\widetilde{g}_{n_y}(0)\rangle_k < 3.16$, which are comparable to the amplitude corresponding to an infinitely thick superconductor. Note that the optimal regime can only be realized for a specific film thickness, $L_z \approx 6.34~$nm.}
\label{FIG3}
\vspace{-1mm}
\end{figure}

Next, we characterize the dependence of the zero-frequency relative amplitude $\widetilde{g}$ on the film thickness for a SC film of finite width $L_y=51.3~$nm, which corresponds to $N_y = 42$ lattice sites. Note that the energy window ($\pm 10~$meV) used  for the averaging in Fig. \ref{FIG2} is somewhat arbitrary. Instead, we focus now on the wave vector range $[0, k_{max}]$ that is relevant for the low-energy physics in SM-SC hybrid structures. Consider, for concreteness, that the semiconductor material is InAs. Assuming that multiple confinement-induced SM bands are occupied, the Fermi k-vector associated with the lowest-energy occupied band is of the order of $0.25~$nm$^{-1}$, which corresponds to a Fermi energy $\sim 100~$meV. This energy scale corresponds to (at most) 4-5 occupied $n_y$ modes.  For a clean system, the relative amplitude associated with the transverse mode $n_y$ and k-vecror $k$ (along the $x$ direction) is
\begin{equation}
\widetilde{g}_{n_y}(\omega, k) = \widetilde{g}(\omega, \epsilon_{n_y} + \epsilon_k),       \label{Eq17}
\end{equation}  
where $\epsilon_{n_y} = 2t(1-\cos[n_y\pi/(N_y+1)])$ and $\epsilon_k =\hbar^2 k^2/2m$. Note that, due to the difference in the effective mass, the SC energy corresponding to $k_{max}$ is $\epsilon_{k_{max}}\approx 2.4~$meV, i.e., much smaller than the corresponding SM energy. Also, we note that Majorana physics is associated with the top occupied band and is characterized by SM energy scales of up to a few meV. In turn, this implies characteristic wave vectors smaller than about  $0.06~$nm$^{-1}$.

\begin{figure}[t]
\begin{center}
\includegraphics[width=0.48\textwidth]{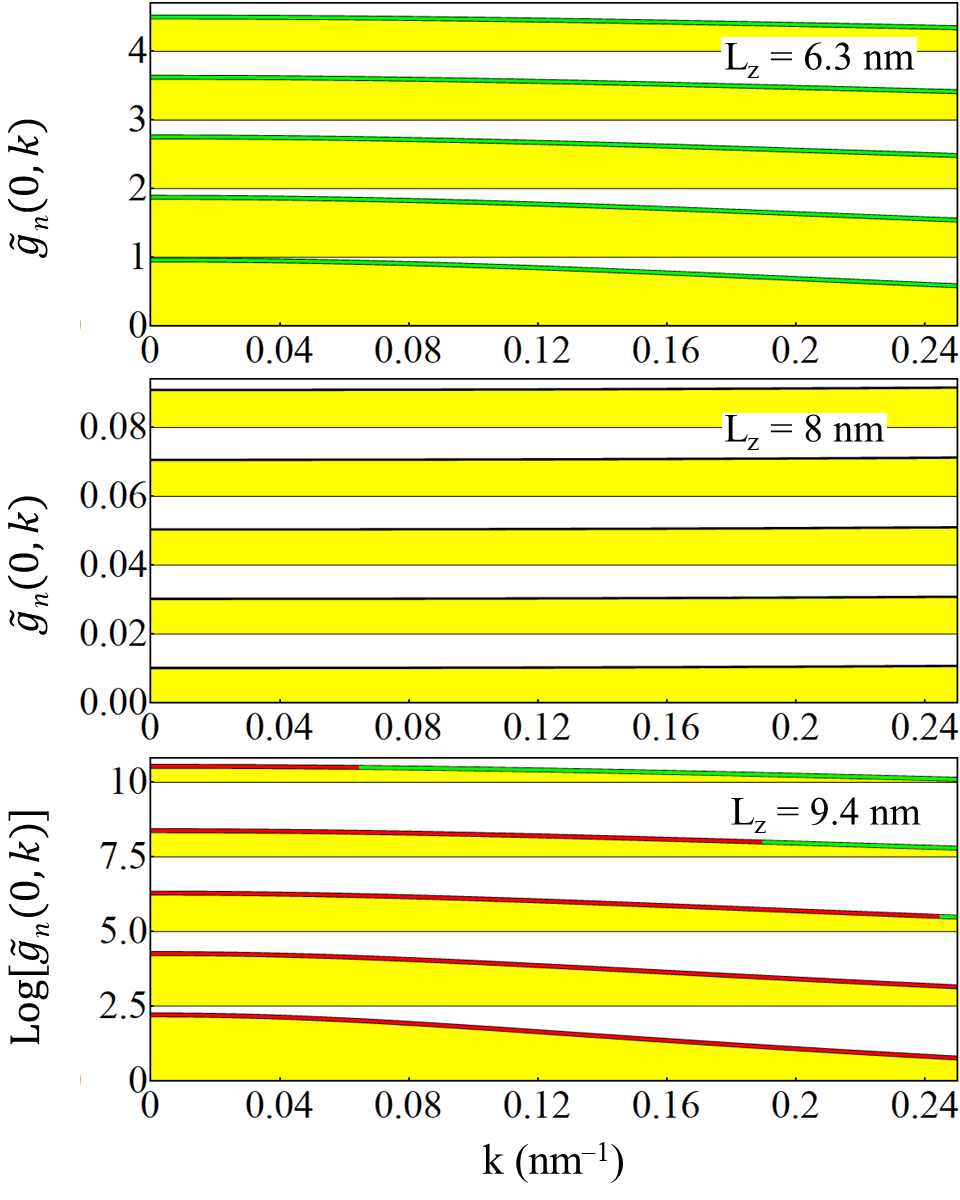}
\end{center}
\vspace{-3mm}
\caption{Dependence of the relative amplitude $\widetilde{g}_{n_y}(\omega, k)$ on the wave vector $k$ (along the $x$ direction) for three different film thicknesses. {\em Top}: optimal regime. Note that $\widetilde{g}_{n_y}(\omega, k)$ has a moderate, smooth dependence on $k$ and $n_y$. The curves corresponding to different transverse modes are shifted by $n_y-1$, for clarity.  {\em Middle}: low $\widetilde{g}$ regime. Note that $\widetilde{g}_{n_y}(\omega, k)$  is almost independent on $k$ and $n_y$. The curves are shifted by $0.02(n_y-1)$. {\em Bottom}: large $\widetilde{g}$ regime.  Note the very strong dependence on $k$ and $n_y$ (by up to two orders of magnitude). The curves are shifted by $2.5(n_y-1)$. The green segments are in the optimal regime ($\widetilde{g}_{n_y} < 3.16$) while red lines correspond to  $\widetilde{g}_{n_y} > 3.16$.}
\label{FIG4}
\vspace{-1mm}
\end{figure}

The thickness dependence of the relative amplitude averaged over $k\in[0,k_{max}]$, $\langle \widetilde{g}_{n_y}(0)\rangle_k$, for the lowest five $n_y$ modes is shown in Fig. \ref{FIG3}. Note that the curves corresponding to different $n_y$ values are shifted, for clarity. 
Also note the semi-quantitative agreement with the results in Fig. \ref{FIG2}(a) and the relatively weak dependence on $n_y$.
We identify an ``optimal regime'' characterized by $\widetilde{g}$ values of order one, specifically $0.316 < \langle\widetilde{g}_{n_y}(0)\rangle_k < 3.16$ (green dots in Fig. \ref{FIG3}). In this regime the proximity effect induced by the thin film is comparable to that induced by a bulk superconductor. Lower values of $\widetilde{g}$ (black dots in Fig. \ref{FIG3}) are likely to generate a weak superconducting proximity effect, while larger values (red dots in Fig. \ref{FIG3}) are likely to put the hybrid system into the strong coupling regime. None of these two  scenarios is optimal for generating robust topological superconductivity.  

To illustrate the dependence of $\widetilde{g}_{n_y}(\omega, k)$ on the wave vector $k$, we focus on three specific film thickness values: $L_z=6.3~$nm, which corresponds to the ``optimal regime'',  $L_z=8~$nm (low $\widetilde{g}$ regime),  and $L_z=9.4~$nm (large $\widetilde{g}$ regime). In the low $\widetilde{g}$ regime (middle panel in Fig. \ref{FIG4}), there 
is almost no dependence on $k$ and $n_y$, but the Green's function amplitude is about two orders of magnitude lower than the corresponding bulk value, 
which makes the realization of a strong-enough proximity effect extremely challenging. By contrast, the large $\widetilde{g}$ regime (bottom panel in Fig. \ref{FIG4}), is characterized by a 
strong  dependence on $k$ and $n_y$. Note that we use ${\rm Log}[\widetilde{g}]$ (instead of $\widetilde{g}$) to conveniently capture this dependence. 
Also note that the $k$ values corresponding to Majorana physics ($k \leq 0.06$) are in the  large $\widetilde{g}$ regime (red lines) for all five $n_y$ bands. 
Finally, the optimal regime (top panel in Fig. \ref{FIG4}) is characterized by a weak dependence on $k$ and $n_y$, but it is only accessible in films of thickness $L_z=6.3~$nm. We emphasize that deviations from this value by one atomic layer automatically put the system into the  low $\widetilde{g}$ regime. Furthermore, as discussed above, changing the values of the model parameters (in particular the lattice constant $c$) can result in small shifts of the $n_z$ modes (see Fig. \ref{FIG2}), which may make the optimal regime only partially accessible (for certain low-$n_y$ modes) or completely  inaccessible within the $L_z$ range considered here. 

We conclude that generating a robust superconducting proximity effect by using a clean, thin SC film would require fine tuning the film thickness with atomic precision and, ultimately, sheer luck. In addition, the presence of an atomic step would automatically result in a highly inhomogeneous hybrid system, with regions of different SC film thickness (e.g., along a proximitized SM wire) being characterized by strengths of the proximity effect that differ by 1-3 orders of magnitude. This behavior is generated by the Fermi surface mismatch problem in combination with the strong finite size quantization of the transverse modes and results in the practical impossibility of inducing a robust, well-controlled proximity effect using clean thin SC films.

\subsection{Superconducting film with surface disorder}\label{S3B}

Our next objective is to investigate the effect of SC disorder on the interface Green's function. Since the presence of disorder breaks translation symmetry and relaxes the requirement of in-plane momentum conservation across the SC-SM interface, one expects a disorder-induced enhancement of the (typically small) values of $\widetilde{g}$. Indeed, using a simplified 2D disorder model consisting of a random onsite potential $V({\bm r}) = V(x, z)$, Ref. \onlinecite{Reeg2018} reports that ``strong surface disorder lead(s) to a small enhancement of the proximity gap''. A disorder-induced enhancement of the induced SC gap is also found in Refs. \onlinecite{Antipov2018,Winkler2019}, based, again, on a simplified 2D random potential model with $V({\bm r}) = V(y, z)$. It is also claimed (but not shown explicitly) that, ``provided it is sufficiently strong, (disorder) removes the non-monotonic dependence on the thickness of the superconducting layer." In Sec. \ref{S3B1} we investigate in detail the random onsite potential model and we find it somewhat unsatisfactory, in the sense that,  for reasonable values of the disorder potential amplitude, this type of disorder does not eliminate the strong dependence of the interface Green's function on the film thickness (i.e., the variation of $\widetilde{g}$ by 1-3 orders of magnitude for changes of $L_z$ by one atomic layer). In Sec. \ref{S3B2} we introduce a 3D model of surface roughness that addresses this issue.

In the presence of disorder the interface Green's function has the general form given by Eq. (\ref{Eq16}). It is convenient to rewrite the matrices $\widetilde{g}$ and $\widetilde{v}$ using the (in-plane) eigenmodes for the clean system,
\begin{eqnarray}
\psi_{n_x n_y}\!({\bm i}) = \!\frac{2}{\sqrt{\!(N_x\!+\!1)(N_y\!+\!1)}}\sin\frac{n_x i_x \pi}{N_x\!+\!1}\sin\frac{n_y i_y \pi}{N_y\!+\!1},~~~~~ \label{Eq18}
\end{eqnarray}
where ${\bm i}=(i_x, i_y)$ labels the position of a lattice site within the $x-y$ plane. The corresponding matrix elements are defined as 
\begin{equation}
\widetilde{g}_{n_x n_y, n_x^\prime n_y^\prime} = \sum_{{\bm i},{\bm j}} \psi_{n_x n_y}\!({\bm i}) ~\!\widetilde{g}_{{\bm i}{\bm j}}~\! \psi_{n_x^\prime n_y^\prime}\!({\bm j}).
\end{equation}
A similar expression defines the matrix elements of $\widetilde{v}$ in the mode basis. Note that $\widetilde{g}_{n_x n_y, n_x^\prime n_y^\prime}$, which is diagonal in the clean limit, has nonzero off-diagonal matrix elements in the presence of disorder. However, to understand the effect of disorder it is enough to focus on the diagonal contributions, which also facilitates the comparison with the clean case. For the diagonal elements we will use the simplified notation $\widetilde{g}_{n_y}(\omega, k_x) \equiv \widetilde{g}_{n_x n_y, n_x n_y}(\omega)$, with $k_x = n_x \pi/(N_x+1)a$. Note that for a clean system in the long wavelength limit this quantity becomes identical with the relative amplitude defined by Eq. (\ref{Eq17}). 

\subsubsection{Onsite random potential model}\label{S3B1}

We first consider a SC film covered with a $2~$nm thick amorphous oxide layer.  Following Refs. \onlinecite{Antipov2018,Winkler2019}, we model the oxide by adding an onsite disorder potential $V_{dis}({\bm i})$ that takes random values in the range $[-U_d, U_d]$ for $i_z c \leq 2~$nm. Note that the disorder potential has zero average and is spatially uncorrelated, 
\begin{equation}
\langle V_{dis}({\bm i}) V_{dis}({\bm j}) \rangle = \frac{U_d^2}{3} ~\!\delta({\bm i}-{\bm j}),
\end{equation}
where $\langle\dots\rangle$ represents averaging over disorder realizations. Also note that the actual strength of the disorder potential depends on both the amplitude $U_d$ and the lattice constants, so that different models correspond to the same disorder strength if the quantity $K_d=ca^2 ~\!U_d^2$ has the same value. Consider, for example, a disorder potential with amplitude $U_d=1~$eV on a lattice with parameters $a=c=0.1~$nm, which corresponds to $K_d = 1000~{\rm meV}^2{\rm nm}^3$, similar to the values used in Refs. \cite{Antipov2018,Winkler2019}. The equivalent disorder for a system with $a=0.38~$nm and $c=0.117~$nm, which correspond to a volume $a^2c$ approximately equal to the volume of the primitive cell of Al, has an amplitude $U_d\approx 243~$meV. We note that having significantly smaller lattice constants (i.e., choosing a finer discretization) would generate a random potential with strong variations on length scales much smaller than the size of the primitive cell, which is rather unphysical. 

\begin{figure}[t]
\begin{center}
\includegraphics[width=0.48\textwidth]{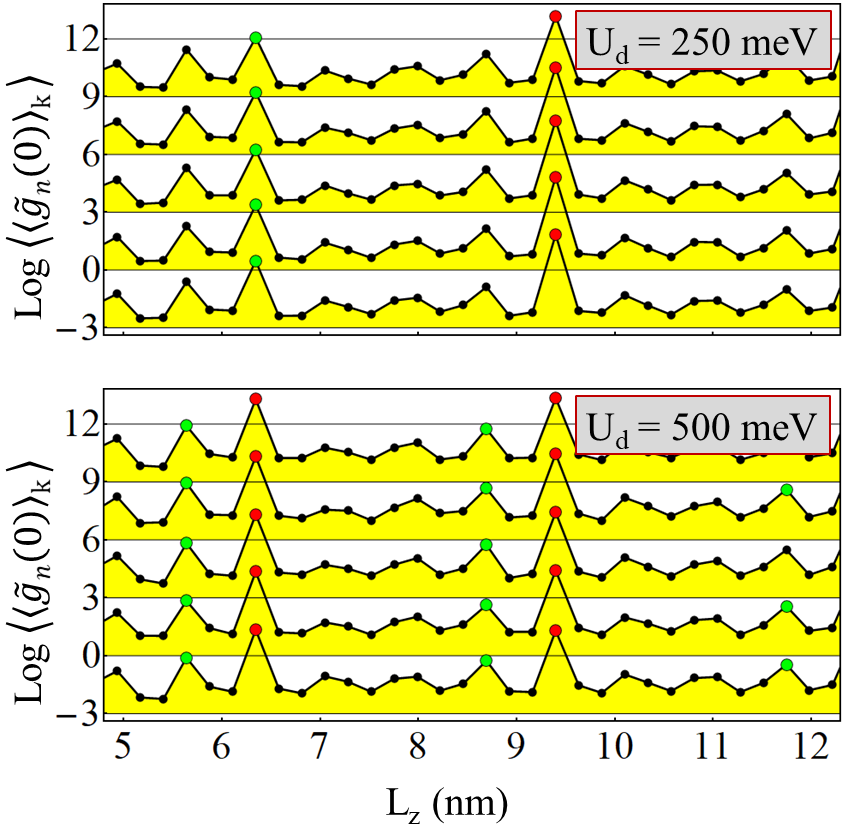}
\end{center}
\vspace{-3mm}
\caption{Thickness dependence of the diagonal relative amplitude $\widetilde{g}_{n_y}(\omega, k)$ averaged over $k\in[0, 0.25~{\rm nm}^{-1}]$ and over 50 different disorder realizations for an infinitely long  SC system with $L_y=50.2~$nm and 2D random potential disorder within $2~$nm from the surface. The film thickness $L_z$ includes the $2~$nm oxide layer. The curves corresponding to the lowest five $n_y$ modes are shifted by $3(n_y-1)$, for clarity. Note the rather weak dependence on $n_y$.
The lattice parameters are $a=0.38~$nm and $c=0.117~$nm, which implies $K_d\approx 1060~{\rm meV}^2{\rm nm}^3$ (top panel) and $K_d\approx 4240~{\rm meV}^2{\rm nm}^3$ (bottom panel). Adding onsite disorder enhances the relative amplitude $\widetilde{g}$ of the interface Green's function (also see Fig. \ref{FIG6}) and slightly alleviates its strong thickness dependence, but does not eliminate it.}
\label{FIG5}
\vspace{-1mm}
\end{figure}

We start with a simplified 2D model that assumes a translation invariant system along the $x$ direction, i.e., we consider $V_{dis}({\bm i}) = V_{dis}(i_y, i_z)$. We calculate the relative amplitude $\widetilde{g}_{n_y}(\omega, k)$ averaged over $k$, with $0\leq k \leq 0.25~$nm$^{-1}$, and over 50 different disorder realizations. The results are shown in Fig. \ref{FIG5}.  Comparison with Fig. \ref{FIG3} (the clean case) clearly shows that the presence of disorder enhances the relative amplitude $\widetilde{g}$ of the interface Green's function and, consequently, the strength of the superconducting proximity effect induced by the thin SC film. Note, however, that despite this enhancement a disorder potential with $K_d\approx 1060~{\rm meV}^2{\rm nm}^3$ (top panel in Fig. \ref{FIG5}), which is similar to the effective disorder strength used in Refs. \cite{Antipov2018,Winkler2019}, does not eliminate the strong dependence of  $\widetilde{g}$ on the film thickness, $L_z$. In particular, one can still notice variations of  $\widetilde{g}$ by up to three orders of magnitude when $L_z$ changes  by one atomic layer. This issue, which we dubbed the ``fundamental problem of proximity-inducing superconductivity using thin, clean SC films'', persists even when we double the amplitude of the random onsite potential to $U_d=500~$meV (bottom  panel in Fig. \ref{FIG5}), although it is manifestly less dramatic than in the clean case (see Fig. \ref{FIG3}). A more detailed analysis of the dependence of the amplitude $\widetilde{g}$ on the disorder potential is provided in Appendix \ref{AppA} (see Fig. \ref{FIGA1}). In addition, the results in App. \ref{AppA} show explicitly that (i) reducing the in-plane lattice constant (to $a=0.285~$nm) while correspondingly increasing the amplitude $U_d$ so that $K_d$ remains constant generates an ``equivalent'' disorder potential (see Fig. \ref{FIGA2}) and (ii) reducing the thickness of the oxide layer (to $1~$nm) slightly reduces the disorder-induced enhancement of $\widetilde{g}$, but does not affect significantly its thickness dependence (Fig. \ref{FIGA3}). Finally, we find that the strong thickness dependence of the interface Green’s function can only be eliminated if we consider extremely strong, unphysical disorder potentials, e.g., $U_d = 10 ~$eV in Fig. \ref{FIGA4}.

\begin{figure}[t]
\begin{center}
\includegraphics[width=0.48\textwidth]{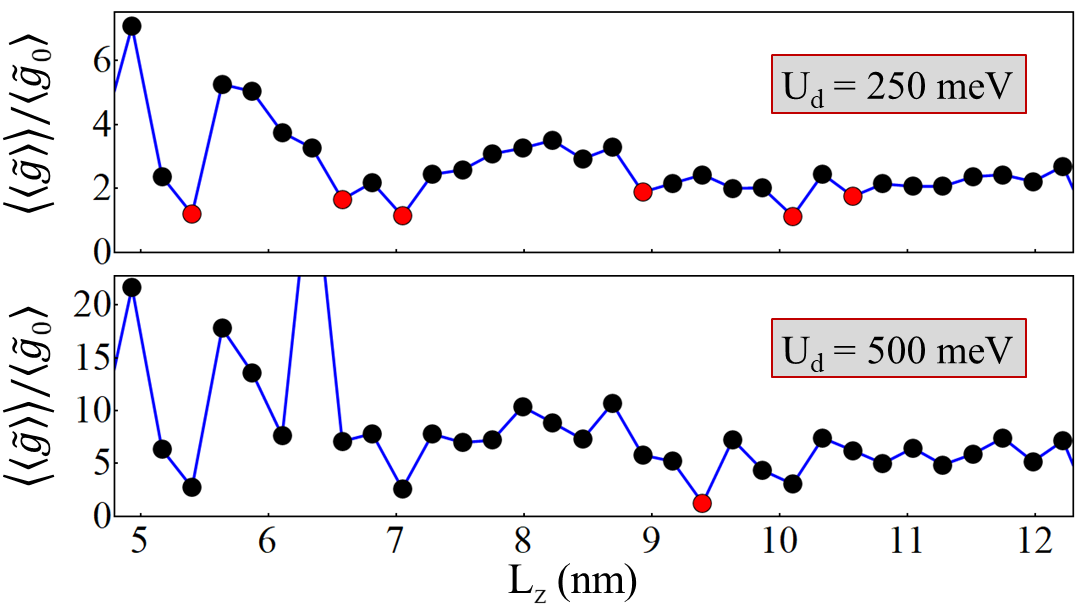}
\end{center}
\vspace{-3mm}
\caption{Enhancement of the average amplitude $\widetilde{g}$ in the presence of disorder. The quantity $\langle\langle\widetilde{g}\rangle \rangle$ is the  average relative amplitude shown in Fig. \ref{FIG5} further averaged over the transverse modes $n_y=\overline{1,5}$, while $\langle\widetilde{g}_0\rangle$ is the relative amplitude of a clean system (with the same dimensions as the corresponding disordered wire) averaged over $k\in[0, 0.25~{\rm nm}^{-1}]$ and  over the transverse modes $n_y=\overline{1,5}$ (also see Fig. \ref{FIG3}). The red dots correspond to disorder-induced enhancements by factors smaller than 2.}
\label{FIG6}
\vspace{-1mm}
\end{figure}

To better quantify the disorder-induced enhancement of the interface Green's function amplitude, we compare the average relative amplitude $\widetilde{g}$ in the presence of disorder with the average relative amplitude $\widetilde{g}$ of the clean system. The parameters for the disordered system are the same as in Fig. \ref{FIG5}. In addition to averaging over $k$ and disorder realizations, we also average $\widetilde{g}$ over the lowest five $n_y$ modes. For the clean system, the parameters are practically the same as in Fig. \ref{FIG3}, except for a slightly different width or the wire, $L_y=50.2~$nm, which coincides with the value in Fig. \ref{FIG5}. Again, in addition to the k-average,  we also average over the lowest five $n_y$ modes. The dependence of the averaged disorder-induced enhancement on the film thickness is shown in Fig. \ref{FIG6}. 
For $U_d=250~$meV, the presence of disorder enhances the relative amplitude of the interface Green function by a factor that typically ranges from  2 to 5 (top panel in Fig. \ref{FIG6}). Stronger disorder,   $U_d=500~$meV (lower panel), results in a larger enhancement of $\widetilde{g}$ (typically by a factor 5-10), except for $L_z = 9.4~$nm, when the clean system is in the large $\widetilde{g}$ regime (see Fig. \ref{FIG3}). We note that, in general, the disorder-induced enhancement is significant, but definitely not ``dramatic'', particularly if we take into account the small typical values of $\widetilde{g}$ (of order $10^{-3}-10^{-1}$) that characterize the clean system (see Fig. \ref{FIG3}).

\begin{figure}[t]
\begin{center}
\includegraphics[width=0.48\textwidth]{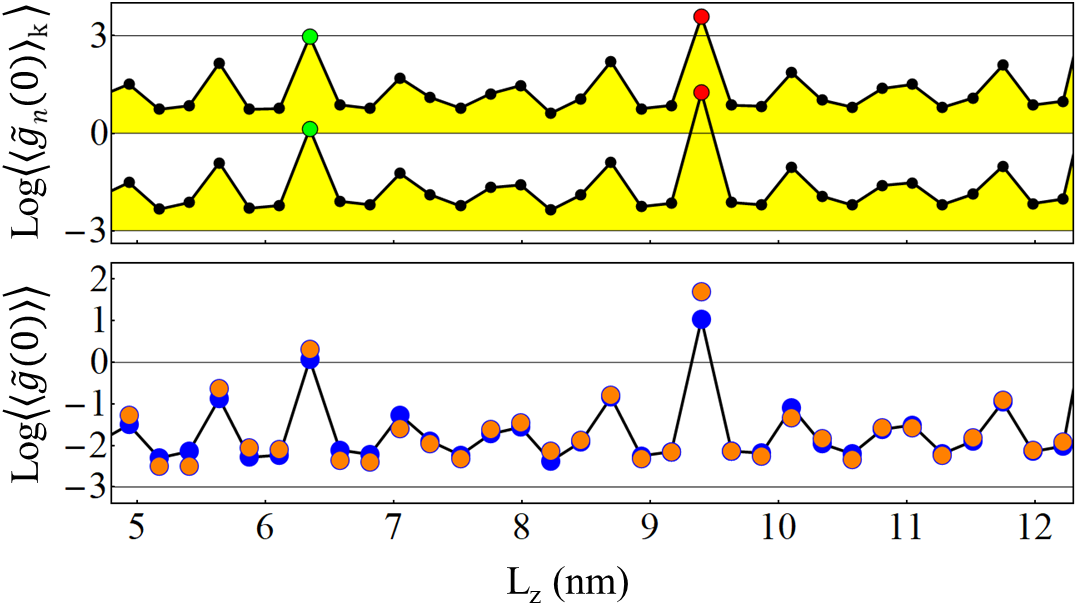}
\end{center}
\vspace{-3mm}
\caption{{\em Top}: Thickness dependence of the average relative amplitude $\widetilde{g}$ for a small finite system in the presence of 3D onsite random potential disorder of amplitude $U_d=250~$meV within $2~$nm from the surface. The finite system has dimensions $L_x=22~$nm and $L_y=20.5~$nm and is defined on a lattice with parameters  $a=0.38~$nm and $c=0.117~$nm, which implies $K_d\approx 1060~{\rm meV}^2{\rm nm}^3$, similar to the top panel of Fig. \ref{FIG5}. The curves corresponding to the lowest two transverse modes, $n_y=1$ and $n_y=2$ (shifted, for clarity), represent the relative amplitude averaged over the lowest two $n_x$ modes and over 10 disorder realizations. {\em Bottom}: Comparison between the average  relative amplitude $\widetilde{g}$ of the system with 3D  random potential disorder (blue dots) and that of the system with 2D  random potential disorder shown in the top panel of  Fig. \ref{FIG5} (orange dots). Note that $\langle\langle\widetilde{g}(0)\rangle\rangle$ is also averaged over $n_y$. }
\label{FIG7}
\vspace{-1mm}
\end{figure}

The natural question is whether the features described above are determined by our use of a simplified 2D disorder model, rather than being generic properties of the (full 3D) onsite random potential model. In particular, is the disorder-induced enhancement affected by the 2D nature of the disorder potential? To address this question,  
we consider now a finite system with $L_x= 22~$nm and $L_y=20.5~$nm having  random onsite disorder of amplitude $U_d=250~$meV within a $2~$nm layer near the surface, i.e. for $1\leq i_z \leq 17$. We calculate the diagonal elements  $g_{n_y}(\omega, k_x)$ averaged over $k_x \lesssim 0.28~{\rm nm}^{-1}$, which implies $n_x \leq 2$, and over 10 different disorder realizations. The results are shown in Fig. \ref{FIG7}. Note the similarities with the results based on the 2D disorder model corresponding to the same effective disorder strength, $K_d\approx 1060~{\rm meV}^2{\rm nm}^3$, which are shown in the top panel of Fig. \ref{FIG5}. In the bottom panel of Fig. \ref{FIG7} we directly compare the diagonal relative amplitudes averaged over $k$ ($k_x$), $n_y$, and disorder realizations for the two disorder models. Based on the semi-quantitative agreement between the two sets of results, we conclude that 2D  random potential disorder  has essentially the same effect on the (average)  interface Green's function amplitude as 3D random potential disorder. We emphasize that the numbers of $n_x$ and $n_y$ modes explicitly considered in the calculation depend on the size of the system trough the condition that the corresponding wave vectors, $k_{x(y)} = n_{x(y)} \pi/L_{x(y)}$, be less than $\sim 0.3~$nm$^{-1}$. We also note that considering a relatively small system ($L_x= 22~$nm, $L_y=20.5~$nm) is physically meaningful because in the presence of disorder the Green's function $\widetilde{G}(\omega; {\bm i},{\bm j})$ becomes short ranged. This important feature will be discussed in detail in the next section.

Our conclusion, so far, is that the presence of disorder typically enhances the amplitude of the interface Green's function and, consequently, the proximity effect induced by the thin SC film. However, adding an onsite disorder potential  does not eliminate the strong dependence of the interface Green's function on the film thickness, $L_z$, which results in variations of $\widetilde{g}$ by up to three orders of magnitude when $L_z$ changes by one atomic layer. We note that extremely large values of the random potential amplitude ($U_d\sim 10~$eV) would solve this issue and would generate an interface Green's function that is practically indistinguishable from that generated by a bulk superconductor (see App. \ref{AppA}), but assuming such large amplitudes is physically problematic. Finally, note that, given a certain disorder strength, the 3D disorder model does not generate more enhancement than the simplified 2D model.   

\subsubsection{Surface roughness}\label{S3B2}

The onsite disorder model investigated in Section \ref{S3B1} is rather unsatisfactory. First, since the surface oxide has a large gap \cite{Mo1998}, one expects the amplitude of the wave function in this region to practically vanish, which, in turn, completely suppresses the effect of the random potential $V_{dis}$.  Most importantly, the model cannot explain the absence of a strong thickness dependence of the interface Green's function, which is associated with the confinement-induced quantization along the $z$ direction. 
Here, we consider a different disorder model based on the assumption that the surface of the superconductor (or the interface with the insulating oxide) has roughness corresponding to variations of $L_z$ by a few atomic layers, e.g., $|\delta L_z| < 0.23-0.35~$nm corresponding to variations by $\pm(1-1.5)$ atomic layers. The technical details associated with the modeling of surface roughness are described in Appendix \ref{AppB}. In essence, we generate a random profile with controlled in-plane and transverse length scales that lays within the top few layers of the thin SC film. The lattice points that are above this profile (i.e., inside the vacuum or oxide region) are made inaccessible by defining a potential larger than the Fermi energy. The potential below the random profile (i.e., inside the superconducting Al) is zero.
This defines a rough surface (with controllable parameters) that will generate electron scattering and mixing of ``clean'' quantum modes. The effects of the rough surface on the interface Green's function are investigated below.

\begin{figure}[t]
\begin{center}
\includegraphics[width=0.48\textwidth]{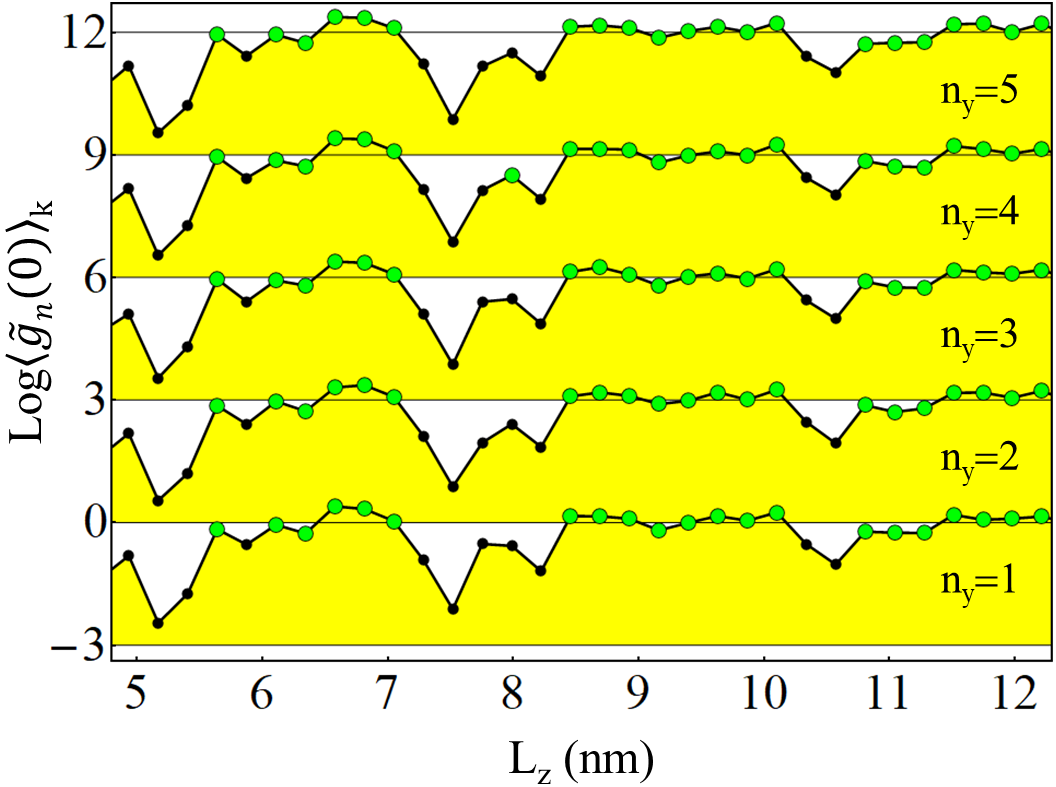}
\end{center}
\vspace{-3mm}
\caption{Dependence of the average relative amplitude, $\langle\widetilde{g}_{n_y}(0)\rangle_{k_x}$, on the SC film thickness, $L_z$, for a  system with surface roughness corresponding to the specific realization shown in Fig. \ref{FIG1}(c). The wire has length $L_x=1.025~\mu$m and width $L_y=51.3~$nm.  The relative amplitude for the lowest five $n_y$ modes  was averaged over $k_x$, with $0 \leq k_x \lesssim 0.25~$nm$^{-1}$, which corresponds to $n_x\leq 82$. The corresponding curves are shifted by $3(n_y-1)$, for clarity. Note the absence of a strong thickness dependence characteristic of the clean case (see Fig. \ref{FIG3}), which was still present in the onsite random potential model (Figs. \ref{FIG5} and \ref{FIG7}). The values of  $\langle\widetilde{g}_{n_y}(0)\rangle_{k_x}$ are in the optimal regime within three relatively wide thickness ranges. The corresponding disorder-induced enhancement of $\tilde{g}$ is dramatic, by up to three orders of magnitude, except for $L_z=9.4~$nm, which corresponds to the large $\widetilde{g}$ regime in a clean system (see Fig. \ref{FIG3}), when the relative amplitude is actually suppressed by the surface disorder.}
\label{FIG8}
\vspace{-1mm}
\end{figure}

We begin with the thickness dependence of the average relative amplitude of the interface Green's function in the presence of surface roughness. In particular, we consider a finite SC film of dimensions $L_x=1.025~\mu$m, $L_y=51.3~$nm and different values of $L_z$ in the presence of the specific surface roughness realization shown in Fig. \ref{FIG1}(c) and we calculate the corresponding relative amplitude $\widetilde{g}_{n_y}(0, k_x)$ averaged over $k_x=n_x\pi/L_x$ as a function of the film thickness $L_z$. The results are shown in Fig. \ref{FIG8}. The striking difference between these results and the clean case in Fig. \ref{FIG3},  or the results in Sec. \ref{S3B1} (see, e.g., Figs. \ref{FIG5} and \ref{FIG7}) is the absence of a strong $L_z$ dependence within significant thickness ranges (e.g., $8.46 \leq L_z \leq 10.1~$nm).  Moreover, within these ranges, the relative amplitude is in the optimal regime, i.e., it has values comparable with those corresponding to a bulk superconductor. This is one of the key results of this study, showing explicitly that surface disorder (in this case surface roughness) can result in the interface Green's function of the thin SC film being practically equivalent to the interface Green's function of a bulk SC over significant thickness ranges. In other words, the presence of such surface disorder solves the ``fundamental  problem''  of proximity-inducing superconductivity using thin SC films. 
We note that the surface roughness used in the calculation was not ``optimized'' to minimize or eliminate the low-$\widetilde{g}$ ``windows'' (black dots in Fig. \ref{FIG8}). 

\begin{figure}[t]
\begin{center}
\includegraphics[width=0.48\textwidth]{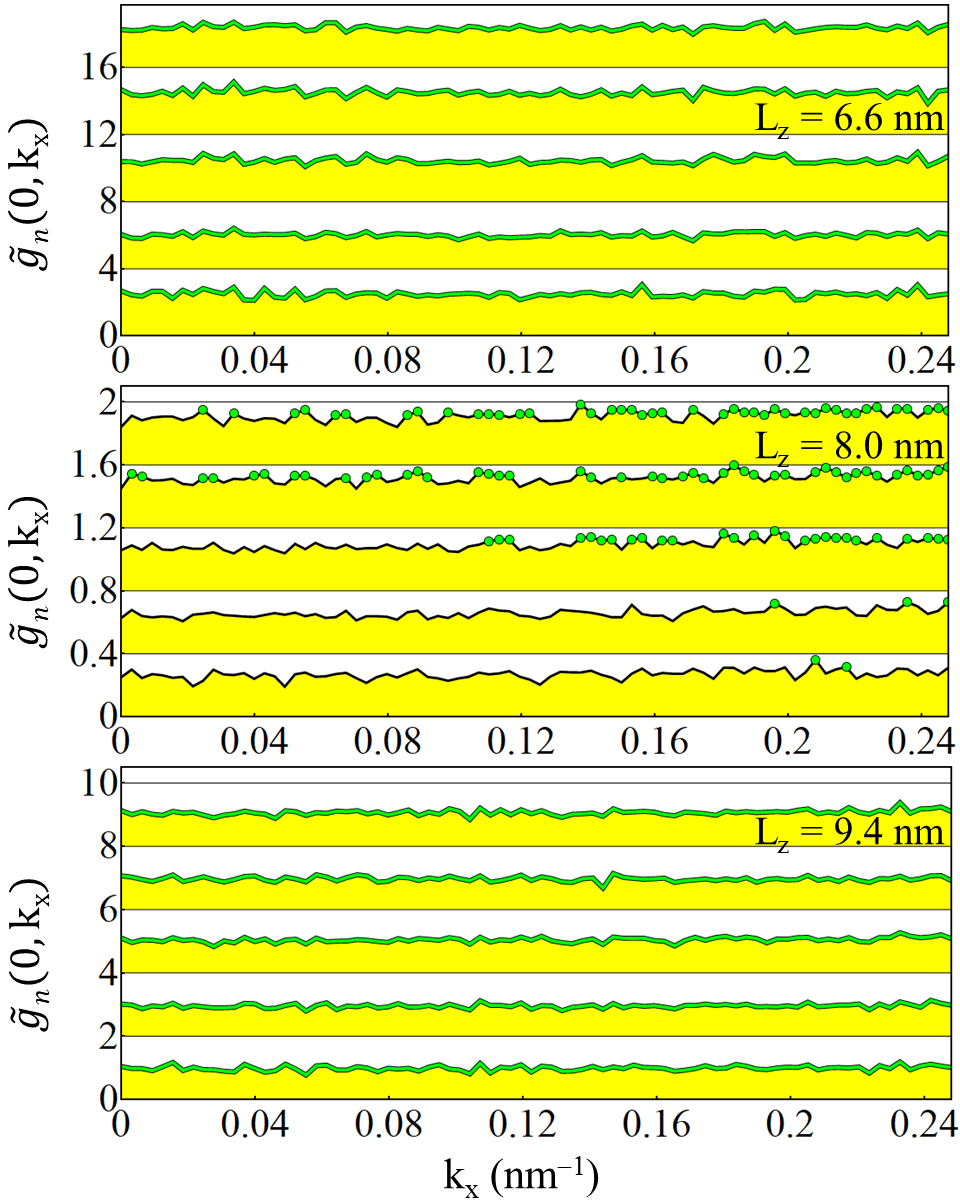}
\end{center}
\vspace{-3mm}
\caption{Dependence of the relative amplitude $\widetilde{g}_{n_y}(\omega, k_x)$ on the ``wave vector'' $k_x=n_x \pi/L_x$  for a system with surface roughness having the same parameters as in Fig. \ref{FIG8} and three different values of the film thickness, $L_z$. Note that $k_x\lesssim 0.25~$nm$^{-1}$ corresponds to the longitudinal modes with $n_x\leq 82$. The dependence on $k_x$ is characterized by small (disorder-induced) fluctuations, but overall it is rather weak.}
\label{FIG9}
\vspace{-1mm}
\end{figure}

For completeness, we also consider the explicit dependence of the relative amplitude on the ``wave vector'' $k_x=n_x\pi/L_x$, which is the correspondent of the dependence shown in Fig. \ref{FIG4} for a clean system. Specifically, Fig. \ref{FIG9} shows the dependence of the (diagonal)  relative amplitude $\widetilde{g}_{n_y}(\omega, k_x)$ on the ``wave vector'' $k_x$ for a system with surface roughness having the same parameters as in Fig. \ref{FIG8}.  Note that, except some small disorder-induced fluctuations, the overall dependence on $k_x$ (i.e., on the longitudinal mode $n_x$) and on $n_y$ is weaker than the dependence shown in the top panel of  Fig. \ref{FIG4}, which corresponds to the ``optimal regime'' of a clean SC film (and strikingly weaker that the dependence shown in the bottom panel of  Fig. \ref{FIG4}).
In particular, for $L_z=6.6~$nm (top panel) and $L_z=9.4~$nm (bottom panel), which are thickness values within the ``optimal ranges'' shown in Fig. \ref{FIG8},  the interface Green's function has roughly the same value for all $n_y$ and $k_x$ modes and, consequently, the strength of the superconducting proximity effect  affecting different low-energy modes in the semiconductor is solely determined by the transverse profile of these modes, i.e., by their amplitude at the SM-SC interface. If, for example, these amplitudes are comparable, the corresponding induced gaps will be comparable, as well.  In other words, in-plane wave vector matching plays no role in the proximity effect induced by the SC thin film with surface roughness. 
As a minor point, we note that surface disorder does not always enhance  $\widetilde{g}$. For $L_z=9.4~$nm, the presence of surface roughness has reduced the value of  $\widetilde{g}_{n_y}(\omega, k_x)$ by up to two orders of magnitude, as the comparison between the lower panels in Figs. \ref{FIG4} and \ref{FIG9} shows. 

\begin{figure}[t]
\begin{center}
\includegraphics[width=0.48\textwidth]{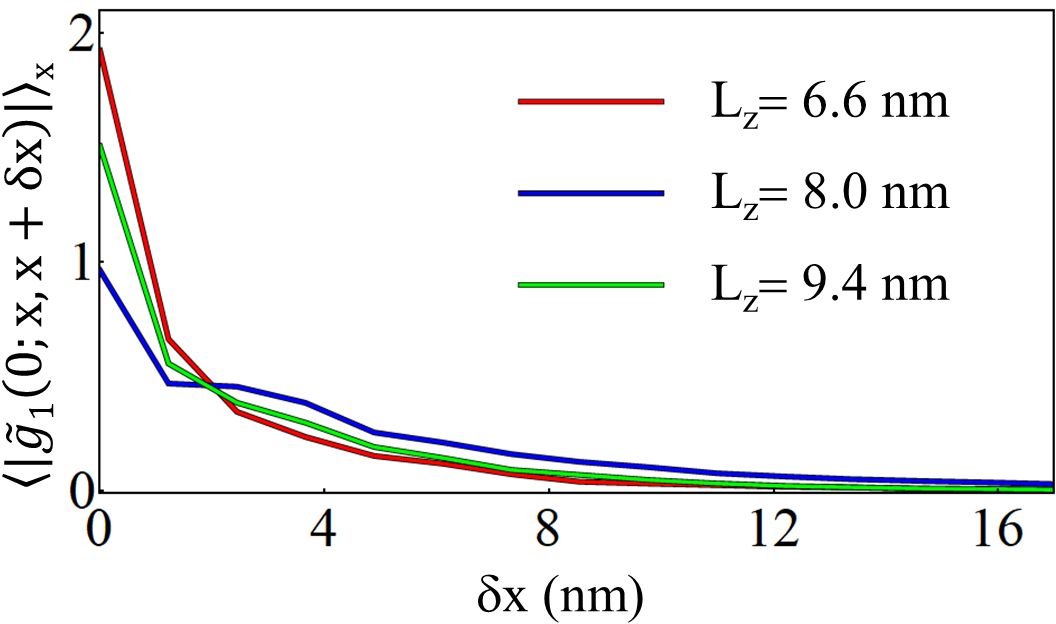}
\end{center}
\vspace{-3mm}
\caption{Dependence of the off-diagonal relative amplitude $|\widetilde{g}_{n_y}(\omega; x, x^\prime)|$ on the longitudinal distance $\delta x = |x^\prime - x|$ between two points. The quantity was averaged over all positions $x$ along the wire.
 Note the sharp decay of  $\widetilde{g}$ with increasing $\delta x$, which reveals that in the presence of surface disorder the interface Green's function becomes short ranged (i.e., quasi-local). Also note that the characteristic decay length is significantly smaller than the width of the wire ($L_y=51.3~$nm). We have $\omega=0$, $n_y=1$, and three different film thicknesses, all other parameters being the same as in Fig. \ref{FIG8}.}
\label{FIG10}
\vspace{-1mm}
\end{figure}

\begin{figure}[t]
\begin{center}
\includegraphics[width=0.48\textwidth]{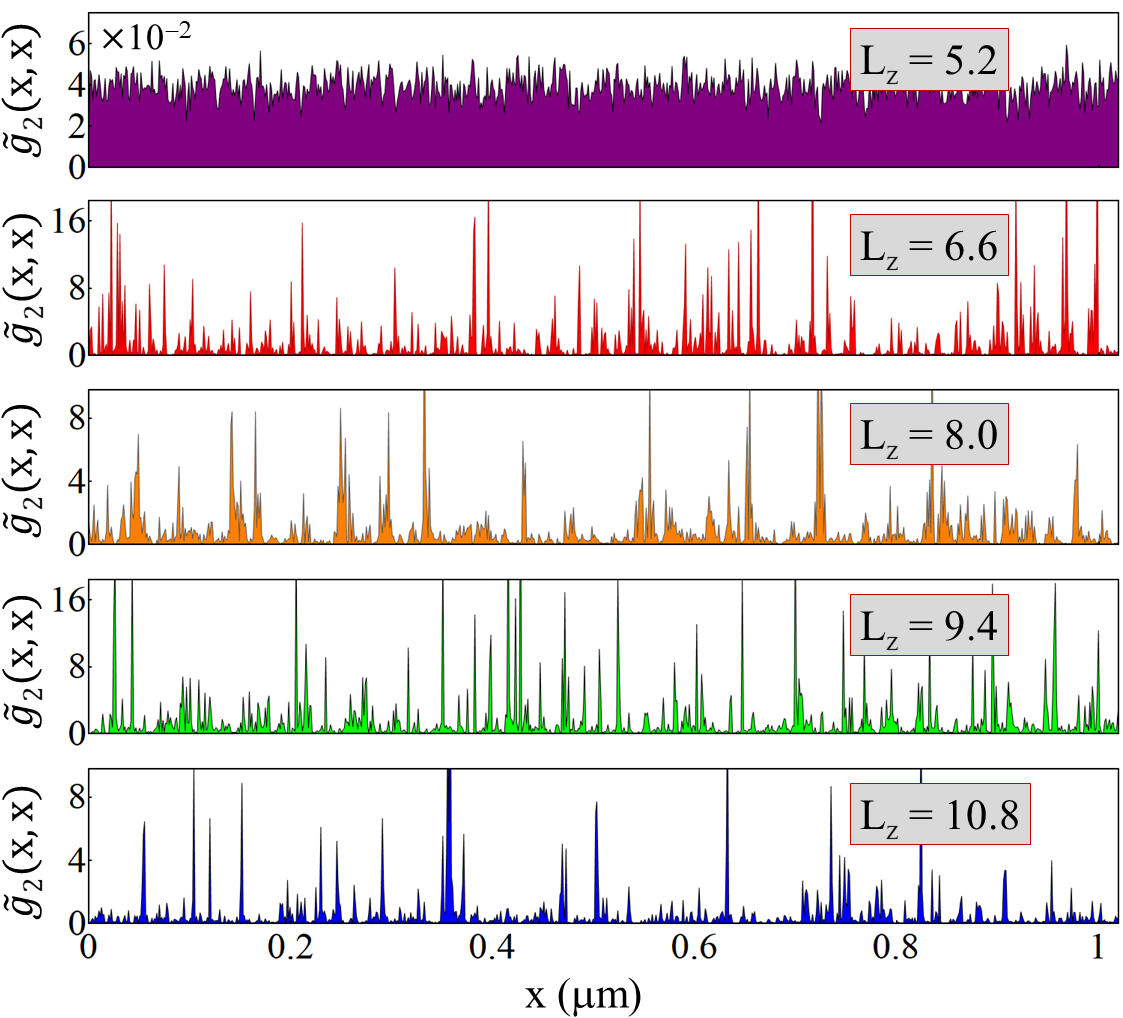}
\end{center}
\vspace{-3mm}
\caption{Onsite relative amplitude $\widetilde{g}_{n_y}(\omega; x, x)$ as a function of the position $x$ along the wire for a system with parameters as in Figs. \ref{FIG8}-\ref{FIG10} and different film thicknesses. Note the different scale in the top panel. For $L_z=5.2~$nm (top panel) the impact of surface roughness is minimal, which is consistent with the corresponding weak disorder-induced enhancement of  $\widetilde{g}$ shown in Fig. \ref{FIG8}. For all other values of $L_z$ we notice huge position-dependent variations of the local relative amplitude (by up to three orders of magnitude). The results correspond to $n_y=2$ and $\omega=0$. A similar behavior characterizes all relevant $n_y$ modes.}
\label{FIG11}
\vspace{-1mm}
\end{figure}

\begin{figure}[t]
\begin{center}
\includegraphics[width=0.48\textwidth]{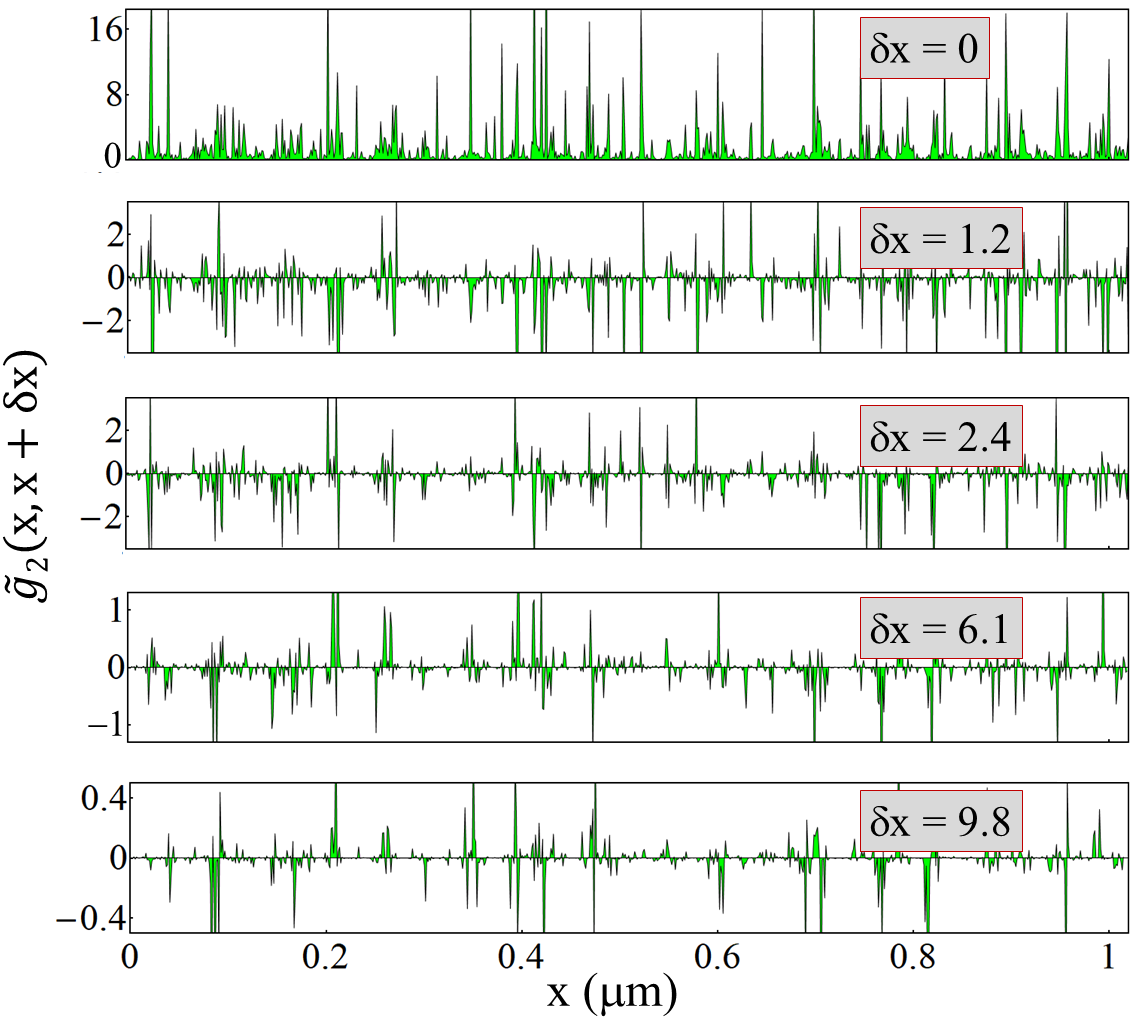}
\end{center}
\vspace{-3mm}
\caption{Relative amplitude $\widetilde{g}_{2}(0; x, x+\delta x)$ as a function of the position $x$ along the wire for a system with parameters as in Figs. \ref{FIG8}-\ref{FIG10} and thickness $L_z = 9.4~$nm.  The values of the local amplitude (top panel, $\delta x = 0$) are the same as in the corresponding panel of Fig. \ref{FIG11}. Note that $\widetilde{g}$ has large position-dependent fluctuations for all values of $\delta x$, but there is an overall decay of the relative amplitude with increasing $\delta x$, consistent with the results in Fig. \ref{FIG10}.}
\label{FIG12}
\vspace{-1mm}
\end{figure}

Next, we switch gears and focus on the real space properties of the relative amplitude $\widetilde{g}(\omega; {\bm i}, {\bm j})$. In particular, we focus on the dependence on the longitudinal variables $x= i_x a$ and $x^\prime = j_x a$ of the matrix elements that are diagonal in the $n_y$ modes,
\begin{equation}
\widetilde{g}_{n_y}(\omega; i_x, j_x) = \sum_{i_y, j_y} \psi_{n_y}(i_y) \widetilde{g}(\omega; {\bm i}, {\bm j}) \psi_{n_y}(j_y),
\end{equation}
where $\psi_{n_y}(i_y)$ is the $y$-component of the eigenmodes of the clean system given by Eq. (\ref{Eq18}).  
We start with the dependence on the longitudinal ``shift'' $\delta x = |x^\prime - x|$, which characterizes the non-locality of the interface Green's function. Fig. \ref{FIG10} shows the dependence of the off-diagonal elements  $\widetilde{g}_{n_y}(\omega; i_x, j_x)$ on the longitudinal distance $\delta x =|i_x-j_x|a$ for the mode $n_y =1$ and three different values of the film thickness. The sharp decay with a characteristic length scale of a few nanometers indicates that, in the presence of disorder, the interface Green's function becomes short ranged. Consequently, we can accurately calculate the (quasi-local) interface Green's function near a certain point $x_0$ by only considering explicitly a certain ``patch'' centered on $x_0$ (rather than simulating the whole system), as long as the ``patch'' is  large-enough as compared to the decay length of the interface Green's function. This is a key result that enables us to simulate large systems by ``covering'' them using multiple patches of numerically manageable size. The details of our patching approach are provided in Appendix \ref{AppC}. Note that the original tight-binding model used in the calculations leading to the results shown in Figs. \ref{FIG8}-\ref{FIG12} is defined on a lattice that has up to $3.67\times 10^6$ points. We efficiently solve this rather challenging numerical problem using the recursive Green's function method (see Sec. \ref{S2}) and the patching approach, which dramatically reduce the numerical cost. 

Now we address the following fundamental question: How is surface disorder ``projected'' onto the interface? In particular, focusing on the low transverse modes $n_y \leq 5$, we investigate the dependence of the  relative amplitude $\widetilde{g}_{n_y}(\omega; x, x+\delta x)$ on the position $x$ along the wire (for fixed values of $\delta x$). We emphasize that addressing this problem requires a 3D disorder model, as simplified 2D models with disorder potentials $V({\bm r})=V(y, z)$ (see, e.g., Sec. \ref{S3B1}) involve translation invariance along the longitudinal ($x$) direction. The dependence of the local (i.e., $\delta x = 0$)  relative amplitude on the position along the wire for systems of various thicknesses is shown in Fig. \ref{FIG11}. For $L_z=5.2~$nm (top panel) the impact of surface roughness is minimal, which is consistent with the result in Fig. \ref{FIG8} showing a weak disorder-induced enhancement of  $\widetilde{g}$ for this value of the film thickness (also compare with Fig. \ref{FIG3}). On the other hand, for all values of $L_z$ that correspond to a significant disorder-induced enhancement (or reduction for $L_z=9.4~$nm) of the relative amplitude we notice large position-dependent variations of  $\widetilde{g}_{2}(0; x, x)$ by up to three orders of magnitude. We note that the other $n_y$ modes are characterized by a similar behavior, but the details (e.g., specific positions and heights of the maxima) are in general different. 

This general behavior also characterizes the off-diagonal components corresponding to $\delta x \neq 0$. The large position-dependent fluctuations characterising the nonlocal relative amplitudes, $\widetilde{g}_{n_y}(\omega; x, x+\delta x)$, for a film of thickness $L_z = 9.4~$nm are illustrated in Fig. \ref{FIG12}.  Note, however, that there is an overall decay of the (average) nonlocal relative amplitude with increasing $\delta x$, consistent with the results in Fig. \ref{FIG10}. We point out that this type of strong position-dependent fluctuations illustrated in Figs. \ref{FIG11} and \ref{FIG12} also characterize the quantity $\widetilde{v}_{n_y}(\omega; x, x+\delta x)$. Hence, we conclude that the presence of surface roughness dramatically enhances the (typically small) interface Green's function of a thin SC film and suppresses its strong thickness dependence over significant $L_z$ windows, but, on the other hand, generates large position-dependent fluctuations of $\widetilde{G}$. The impact of these fluctuations on the stability of induced topological superconductivity (and the correspondent Majorana zero modes) will be addressed in the next section. We note that these key properties of the interface Green's function in the presence of surface disorder are reinforced by the additional results presented in App. \ref{AppD}, which correspond to a different surface roughness realization.

\begin{figure}[t]
\begin{center}
\includegraphics[width=0.48\textwidth]{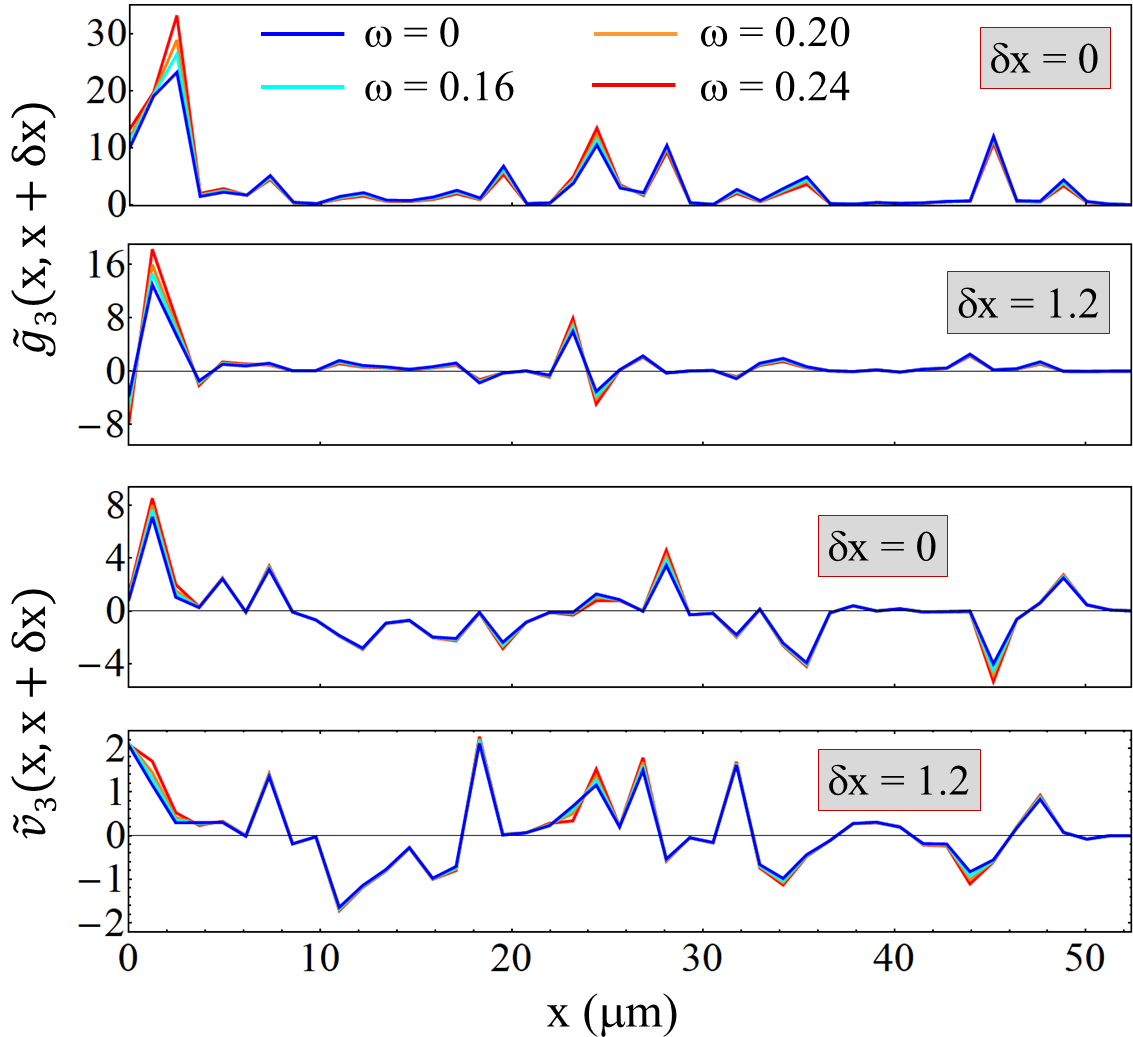}
\end{center}
\vspace{-3mm}
\caption{{\em Top panels}: Relative amplitude $\widetilde{g}_{3}(\omega; x, x+\delta x)$ as a function of position $x$ for two values of $\delta x$ and four different frequencies. {Bottom panels}: Same dependence for the quantity  $\widetilde{v}_{3}(\omega; x, x+\delta x)$. Note that the frequency dependence is negligible for $\omega \lesssim \Delta_0/2\approx 0.16~$meV. The system is a patch of the wire with surface roughness shown in Fig. \ref{FIG1}(c) of dimensions $\ell_x \times  \ell_y = 58.6 \times 51.3~$nm.}
\label{FIG13}
\vspace{-1mm}
\end{figure}

 So far, we have characterized the dependence of the relative amplitude $\widetilde{g}_{{\bm i}{\bm j}}(\omega)$ of the interface Green's function defined by Eq. (\ref{Eq16}), or its representations in terms of in-plane eigenmodes $(n_x, n_y)$ [see Eq. (\ref{Eq18})] on position and/or eigenmode index. It is worth noting again that the quantity $\widetilde{v}_{{\bm i}{\bm j}}(\omega)$, which describes the position and frequency 
dependence of the second term in Eq. (\ref{Eq16}), shares with $\widetilde{g}$ the key properties discussed above. In particular, for a clean system the 
(average)  $\widetilde{v}$ values depend strongly on the film thickness due to the confinement-induced quantization of the $n_z$ modes. This is expected based on our previous discussion of $\widetilde{g}$ and a comparison of the expressions for $\gamma$ and $\zeta$ (which are proportional to  $\widetilde{g}$ and  $\widetilde{v}$, respectively) given by Eqs. (\ref{Eq8}) and  (\ref{Eq9}). Furthermore, including surface disorder alleviates or eliminates this strong thickness dependence, but generates a strong dependence on the position along the wire, as discussed in the context of Figs. \ref{FIG11} and \ref{FIG12}. The remaining question concerns the frequency dependence of both $\widetilde{g}$ and $\widetilde{v}$. To address this question, we consider a patch of the SC film with surface roughness of dimensions  $\ell_x \times  \ell_y = 58.6 \times 51.3~$nm  and calculate the position dependence of  $\widetilde{g}_{n_y}(\omega; x, x+\delta x)$ and  $\widetilde{v}_{n_y}(\omega; x, x+\delta x)$ for different values of $\omega <\Delta_0$. The results are shown in Fig. \ref{FIG13}. 
Remarkably, the dependence on frequency is very weak as long as $\omega$ is not too close to the superconducting gap edge and is practically negligible for $\omega \lesssim \Delta_0/2$. 
In general, this is not the case for a clean system, except in the the large $L_z$ (bulk) limit, when $\widetilde{g}$ and $\widetilde{v}$ become frequency-independent, as discussed in Sec. \ref{S2}.
The quasi-independence of the functions $\widetilde{g}_{{\bm i}{\bm j}}(\omega)$ and $\widetilde{v}_{{\bm i}{\bm j}}(\omega)$ on the frequency (for values of $\omega$ not too close to the gap edge) illustrated in Fig. \ref{FIG13} strengthens our observation that, in the presence of surface roughness, the interface Green's function of a thin SC film is practically equivalent to the corresponding quantity associated with a bulk superconductor, except for the disorder-induced position-dependent fluctuations.

\section{Proximity-induced disorder}\label{S4}

The analysis presented in the previous section has revealed that the presence of surface disorder, in particular surface roughness, has complementary effects on the properties of interface Green's function. On the one hand, it reduces or eliminates the strong dependence on the film thickness and generates an interface Green's function with properties similar to those characterizing a bulk superconductor. In other words, it solves the ``fundamental problem'' of inducing superconductivity using a thin SC film. On the other hand, the presence of surface disorder is ``projected'' at the interface generating strong position-dependent fluctuations of the  Green's function. The next natural question concerns the effect of these fluctuations on the superconductivity proximity-induced by the thin film. What is the effect of the proximity-induced effective disorder generated by the SC film? For concreteness, we focus on a Majorana-type hybrid wire consisting of a semiconductor (SM) nanowire proximity coupled to a thin SC film [see, e.g., Fig. \ref{FIG1}(b)]. The key quantity describing the low-energy properties of the hybrid system is the semiconductor Green's function
\begin{equation}
G_{SM}(\omega) =\left[\omega - H_{SM} -\Sigma_{SC}(\omega)\right]^{-1}, \label{Eq22}
\end{equation}
where $H_{SM}$ is the Hamiltonian describing the SM subsystem. For simplicity, we assume that the inter-band spacing between the relevant transverse SM bands is large, so that we can work within the independent band approximation \cite{Stanescu2017}. In this limit, the relevant low-energy physics is associated with a single band, which can be modeled using the 1D effective Hamiltonian
\begin{eqnarray}
H_{SM} &=& \sum_{i, \sigma}\left[-t_{sm}(\hat{a}_{i\sigma}^\dagger\hat{a}_{i+1\sigma}^{} + {\rm h}.{\rm c}.)+(\epsilon_0 +\mu) \hat{a}_{i\sigma}^\dagger\hat{a}_{i\sigma}^{}\right] \nonumber \\
&+& \frac{\alpha}{2}\sum_{i}\left[(\hat{a}_{i\uparrow}^\dagger\hat{a}_{{i+1\downarrow}}^{} - \hat{a}_{i\downarrow}^\dagger\hat{a}_{{i+1\uparrow}}^{})+ {\rm h}.{\rm c}.\right] \\
&+&\Gamma\sum_{i}(\hat{a}_{i\uparrow}^\dagger\hat{a}_{{i\downarrow}}^{} + {\rm h}.{\rm c}.), \nonumber
\end{eqnarray}
where $t_{sm}$ is the amplitude of the nearest-neighbor hopping, $\epsilon_0=2t_{sm}$, $\mu$ is the chemical potential,  $\alpha$ is the Rashba spin-orbit coupling parameter, and $\Gamma$ is the (half) Zeeman splitting generated by an external field applied along the direction of the  wire. The model is defined on a 1D lattice with lattice constant $a=1.22~$nm (same as the in-plane lattice constant of the SC model) containing $N_x=840$ points, which corresponds to a wire length $L_x=1.025~\mu$m. The values of the model parameters were chosen to be relevant for InAs-based structures: $t_{sm}=1.11~$eV (which corresponds to an effective mass $m^*=0.023m_0$) and  $\alpha\cdot a =250~$meV$\cdot$\AA . 

The last term in Eq. (\ref{Eq22}) is a self-energy contribution defined at the SM-SC interface and describing the proximity effect induced by the SC film after ``integrating out'' the degrees of freedom associated with the superconductor. This self-energy is proportional to the interface Green's function of the SC film discussed in the previous sections.  For simplicity, let us assume that the relevant SM band only couples significantly to one $n_y$ mode, say $n_y=2$. Then, considering the spin and particle-hole structure of the interface Green's function discussed in Sec. \ref{S2}, particularly Eq. (\ref{Eq16}), we have the (first quantized) expression
\begin{eqnarray}
\Sigma_{SC}(\omega) &=&- \Lambda\sum_{i,j} \frac{\widetilde{g}_2(\omega; i,j)}{\sqrt{\Delta_0^2-\omega^2}}\left(\omega ~\!\sigma_0\tau_0+\Delta_0 ~\!\sigma_y\tau_y\right) \nonumber \\
&+&  \Lambda\sum_{i,j} \widetilde{v}_2(\omega; i,j)~\! \sigma_0\tau_z,  \label{Eq24}
\end{eqnarray} 
where $\sigma_\mu$ and $\tau_\mu$ are Pauli matrices associated with the spin and particle-hole degrees of freedom, respectively, and $\Lambda=\pi \nu_{_F}^\infty \widetilde{t}^2$, with $\widetilde{t}$ being an effective hopping across the SM-SC interface. Based on our analysis in Sec. \ref{S3B2}, the frequency dependence of $\widetilde{g}$ and $\widetilde{v}$ is weak and, since we are focusing on the low-energy physics, we can safely neglect it and use the values of these quantities at $\omega=0$. 
 We point out that the term proportional to $\Delta_0$ is responsible for the proximity-induced pairing, while the term proportional to the frequency $\omega$ is responsible for the proximity-induced renormalization of all low-energy quantities \cite{Stanescu2017}. Both these crucial terms are controlled by the relative amplitude $\widetilde{g}_2(i,j)\equiv \widetilde{g}_2(0; i,j)$. In addition, the diagonal contributions proportional to $\widetilde{v}_2(i,i)$ can be viewed as a proximity-induced random potential, while the off-diagonal terms represent (random) hopping matrix elements. 

\begin{figure}[t]
\begin{center}
\includegraphics[width=0.48\textwidth]{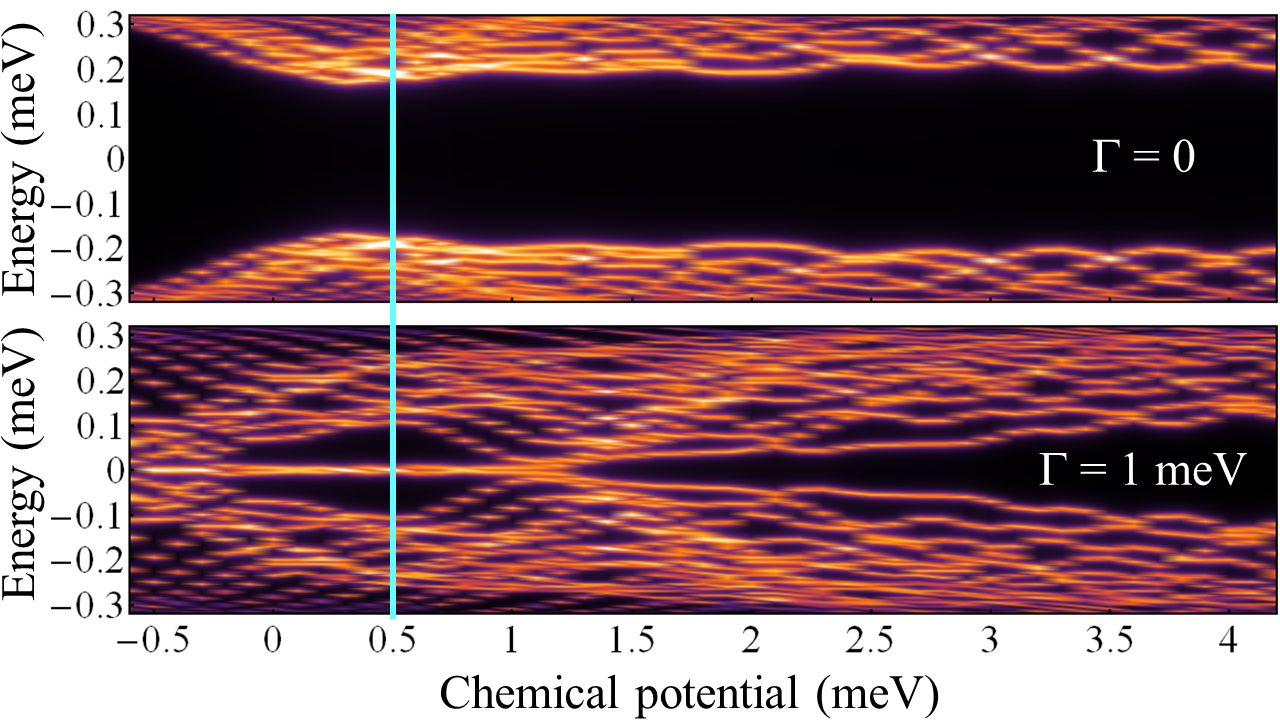}
\end{center}
\vspace{-3mm}
\caption{Dependence of the density of states $\rho(\omega)$ on the chemical potential and frequency/energy for two values of the Zeeman splitting $\Gamma$. The effective SM-SC coupling  is $\widetilde{t}=38.7~$meV ($\Lambda\approx 0.2~$meV). For $\Gamma=0$ (top panel) a finite gap of order $0.2~$meV extends throughout the entire $\mu$ range. For $\Gamma=1~$meV the gap closes and a near-zero energy mode emerges for $\mu\lesssim 1~$meV. The full dependence on the Zeeman field for $\mu=0.5~$meV (vertical cut) is shown in Fig. \ref{FIG15}}
\label{FIG14}
\vspace{-1mm}
\end{figure}

 We emphasize that the effective coupling $\Lambda$ depends on materials properties of the SC (through the density of states $\nu_{_F}^\infty$) and on the transparency of the SM-SC interface combined with the spatial profile of the relevant SM mode, in particular its amplitude at the interface (through the effective hopping $\widetilde{t}$). On the other hand, all properties derived from the thin-film nature of the SC, including the disorder effects, are contained in  $\widetilde{g}$ and $\widetilde{v}$. Without disorder, these quantities (and particularly $\widetilde{g}$) are typically small, which practically implies the absence of induced superconductivity. Disorder (e.g., on the SC surface) enhances these quantities, but also induces effective disorder inside the SM wire (through $\Sigma_{SC}$). To evaluate the effects of the induced disorder, we calculate the SM Green's function given by Eq. \ref{Eq22} with a self energy corresponding to the SC interface Green's function discussed in Sec. \ref{S3B2} for a wire of thickness $L_z=6.6~$nm. Note that the corresponding onsite relative amplitude  $\widetilde{g}_2(i,i)$ is given by the red curve in Fig. \ref{FIG11}. Finally, we calculate the density of states (DOS) or local the local density of states (LDOS) using the relations
\begin{eqnarray}
\rho(\omega) &=& -\frac{1}{\pi}{\rm Im}~\!{\rm Tr}\left[ G_{SM}^{}(\omega+i\eta)\right],  \\
\rho_{_L}(\omega,i) &=& -\frac{1}{\pi}{\rm Im}~\!{\rm Tr}_{_L}\left[ G_{SM}^{}(\omega+i\eta)\right]_{ii},
\end{eqnarray} 
where ${\rm Tr}$ is the trace over position, spin, and the particle-hole degree of freedom, while ${\rm Tr}_{_L}$ is the ``local'' trace over spin and particle-hole variables. Note that both $\rho$ and $\rho_L$ also depend on the control parameters $\mu$ (chemical potential) and $\Gamma$ (Zeeman splitting), as well as on the effective SM-SC coupling ($\widetilde{t}$ or $\Lambda$). 

We start with a calculation of the DOS as function of chemical potential and energy for a system with and without Zeeman field. The results are shown in Fig. \ref{FIG14} for   $\Gamma=0$ (top panel) and $\Gamma=1~$meV (bottom panel). Note the clean induced gap of order $0.6\Delta_0$  characterizing the system in the absence of a Zeeman field, i.e., for $\Gamma=0$. The size of the induced gap indicates that the SM-SC coupling ($\widetilde{t}=38.7~$meV) is in the intermediate regime. Indeed, $\widetilde{t}=38.7~$meV corresponds to $\Lambda\approx 0.2~$meV, while Fig. \ref{FIG9} (top panel) reveals that the low-$k_x$ value of $\widetilde{g}(0,k_x)$ is of order 2; this implies a ``total'' SM-SC effective coupling $\Lambda~\! \widetilde{g}\approx 0.4~$meV, which is comparable to the SC gap of the parent superconductor ($\Delta_0 = 0.33~$meV) placing the hybrid system in the intermediate coupling regime. 
For $\Gamma=1~$meV (bottom panel in Fig. \ref{FIG14}) the gap closes and a near-zero energy mode emerges within a finite $\mu$ range near the bottom of the band. The low-energy  mode is separated from the rest of the spectrum by a finite quasiparticle gap.  This behavior is consistent with the emergence of induced topological superconductivity and Majorana bound states at finite Zeeman field.

\begin{figure}[t]
\begin{center}
\includegraphics[width=0.48\textwidth]{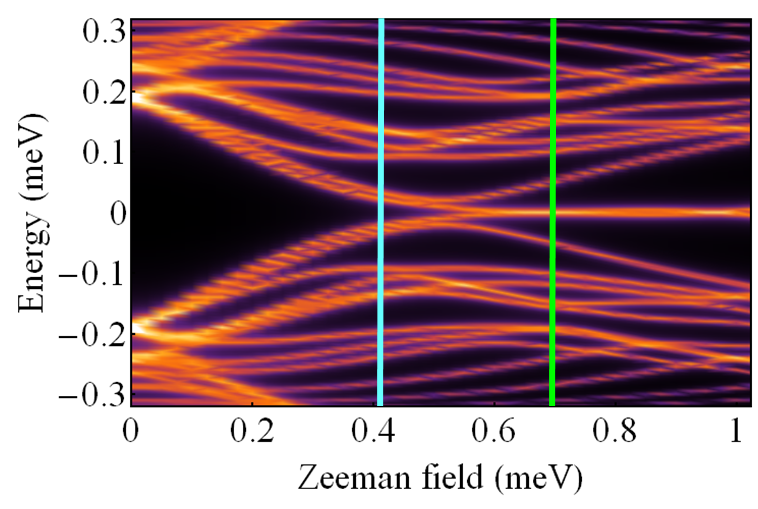}
\end{center}
\vspace{-3mm}
\caption{DOS as a function of Zeeman field and energy for a system with chemical potential $\mu=0.5~$meV and  effective SM-SC coupling  $\widetilde{t}=38.7~$meV. With increasing $\Gamma$, the lowest energy modes merge into a near-zero-energy mode separated from the rest of the spectrum by a finite quasiparticle gap. The LDOS corresponding to $\Gamma=0.4~$meV and $\Gamma=0.7~$meV (vertical cuts) is shown in Fig. \ref{FIG16}.}
\label{FIG15}
\vspace{-1mm}
\end{figure}

Next, we fix the chemical potential at $\mu=0.5~$meV (vertical cut in Fig. \ref{FIG14}) and determine the DOS as function of Zeeman field and energy. The spectrum shown in Fig. \ref{FIG15} has general features that are consistent with those expected for a finite (relatively short) system supporting topological superconductivity and Majorana physics. Note that the type of behavior illustrated in Figs. \ref{FIG14} and \ref{FIG15} represents a generic characteristic of the (minimal) Majorana model in the presence of weak (effective) disorder. 
This result may appear surprising if we consider the large position-dependent fluctuations of the relative amplitude $\widetilde{g}_2$ shown in Fig. \ref{FIG11} (red curve corresponding to $L_z=6.6~$nm). However, we should take into account the characteristic length scale of these fluctuations, which is significantly shorter than the characteristic length scale associated with Majorana physics. The induced disorder gets ``averaged'' over the relevant length scale resulting in a weak effective disorder \cite{Ahn2021}. Further indication that the effective disorder is weaker than suggested by the real space dependence of $\widetilde{g}$ is provided by the ``Fourier transform'' shown in Fig. \ref{FIG9} (top panel), which is characterized by small fluctuations within the relevant $k_x$ range.

\begin{figure}[t]
\begin{center}
\includegraphics[width=0.48\textwidth]{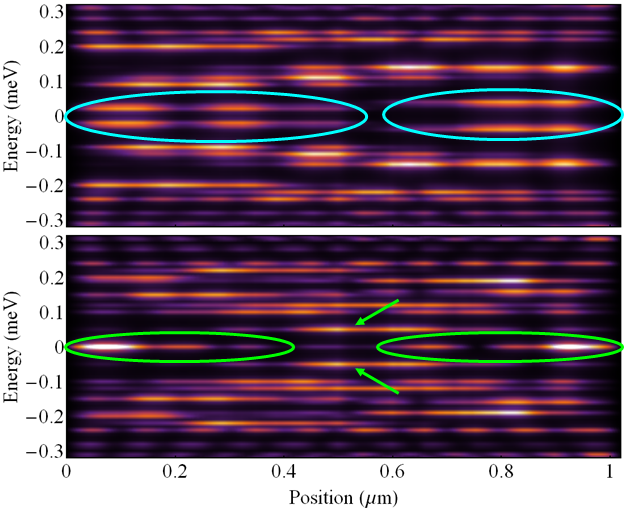}
\end{center}
\vspace{-3mm}
\caption{LDOS as a function of the position along the wire and energy for a system with chemical potential $\mu=0.5~$meV, effective SM-SC coupling $\widetilde{t}=38.7~$meV, and Zeman field $\Gamma=0.4~$meV (top) or $\Gamma=0.7~$meV (bottom). The highlighted features are generated by the lowest energy modes shown in Fig. \ref{FIG15}. In the top panel, these features correspond to two ABSs having slightly different energies and being localized within opposite sides of the wire. In the bottom panel we notice a pair of Majorana modes localized near the opposite ends of the wire and a low-energy ABS localized near the center of the wire (arrows). }
\label{FIG16}
\vspace{-1mm}
\end{figure}

To better understand the effects of weak effective disorder we calculate the LDOS at two representative values of the Zeeman field (marked by vertical cuts in Fig. \ref{FIG15}). The results are shown in Fig. \ref{FIG16}. Let us focus on the features associated with the two lowest energy modes shown in Fig. \ref{FIG15}. Note that, in the Majorana basis, these modes correspond to two pairs of Majoranas (i.e., four Majorana modes).  For $\Gamma \lesssim 0.5~$meV these modes collapse toward zero energy as we increase the Zeeman field. Based on the results shown in the top panel of Fig. \ref{FIG16}, the lowest energy mode is an Andreev bound state (i.e. a pair of overlapping Majorana modes) localized within the left half of the wire, while the second-lowest mode is an ABS localized on the right side of the wire. For $\Gamma \gtrsim 0.5~$meV one mode sticks to zero energy, while the other acquires a gap (see Fig. \ref{FIG15}). As shown in the lower panel of Fig. \ref{FIG16}, the (near) zero energy mode consists of a pair of well-separated Majoranas localized near the opposite ends of the wire, while the finite energy mode is an ABS (i.e., a pair of overlapping Majoranas) localized near the center of the wire. 
We point out that the presence of weak disorder results in the asymmetry of the ABS modes in the top panel (as well as other asymmetric features in both panels) and in the localization of the low-energy ABS in the lower panel.  We emphasize that, as a result of weak disorder, the lowest finite-energy mode (at finite Zeeman field) is not a ``bulk'' state and does not couple significantly to probes attached to the wire ends (particularly the left end). For large-enough values of $\Gamma$ the well-separated Majorana modes are protected by a finite quasiparticle gap. However, we should remind the reader that we have not included effects associated with the presence of an external magnetic field in the calculation of the interface Green's function for the SC film. These effects are expected to suppress the parent SC gap and, implicitly, the induced pairing, eventually driving the system into the strong SM-SC coupling regime. As shown below, this will enhance the impact of induced disorder.

\begin{figure}[t]
\begin{center}
\includegraphics[width=0.48\textwidth]{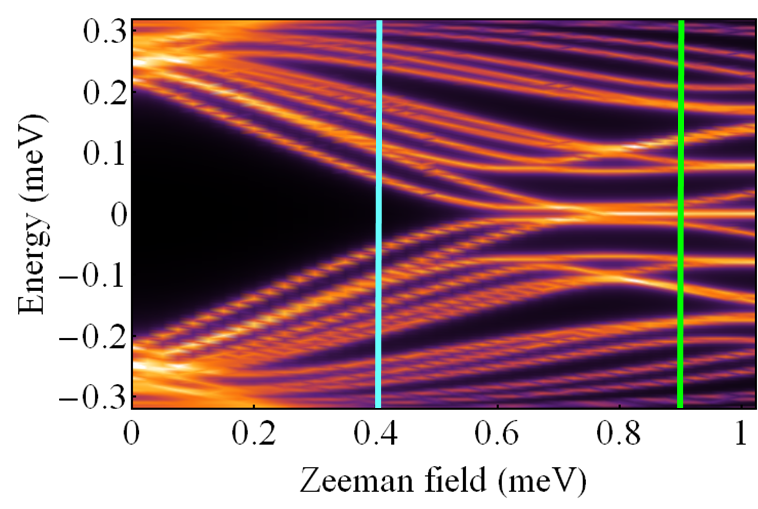}
\end{center}
\vspace{-3mm}
\caption{DOS as a function of Zeeman field and energy for a system with chemical potential $\mu=0.5~$meV and  effective SM-SC coupling  $\widetilde{t}=50~$meV, which corresponds to $\Lambda\approx 0.345~$meV. With increasing $\Gamma$, the lowest energy modes merge into a zero-energy mode and a low-gap finite energy mode. The LDOS corresponding to $\Gamma=0.4~$meV and $\Gamma=0.9~$meV (vertical cuts) is shown in Fig. \ref{FIG18}.}
\label{FIG17}
\vspace{-1mm}
\end{figure}

Since the self-energy given by Eq. \ref{Eq24} is proportional to $\Lambda$, increasing the SM-SC coupling is expected to enhance the induced disorder. To verify this intuition, we consider a stronger SM-SC coupling corresponding to  $\widetilde{t}=50~$meV, i.e., $\Lambda\approx 0.345~$meV. Note that the ``total'' effective coupling is now $\Lambda~\! \widetilde{g}\approx 0.7~{\rm meV}\sim 2\Delta_0$.
The dependence of the corresponding DOS on Zeeman field and energy for a fixed chemical potential $\mu=0.5~$meV is shown in Fig. \ref{FIG17}. First, comparison with Fig. \ref{FIG15} shows that, as a result of increasing the SM-SC effective coupling,  the zero-field induced gap is slightly larger and the collapse toward zero energy of the lowest energy modes occurs at a slightly higher value of the Zeeman field. However, the striking difference is that, after collapsing toward zero energy, the second lowest mode does not acquire a gap larger than the typical inter-state energy spacing. In other words, the near-zero energy mode that emerges above $\Gamma \approx 0.7~$meV is no longer protected by a significant quasiparticle gap. This already suggests that the stability of the Majorana modes may be affected by disorder-induced low-energy states. We note that, in practice, increasing the SM-SC coupling from 
$\widetilde{t}=38.7~$meV to $\widetilde{t}=50~$meV can be easily achieved using a back-gate potential that ``pushes'' the electrons toward the SM-SC interface (recall that $\widetilde{t}$ is roughly proportional to the amplitude of the SM wave function at the interface). This raises a serious concern regarding the practical possibility of simultaneously controlling the chemical potential (of the SM wire) and the strength of the induced disorder.

\begin{figure}[t]
\begin{center}
\includegraphics[width=0.48\textwidth]{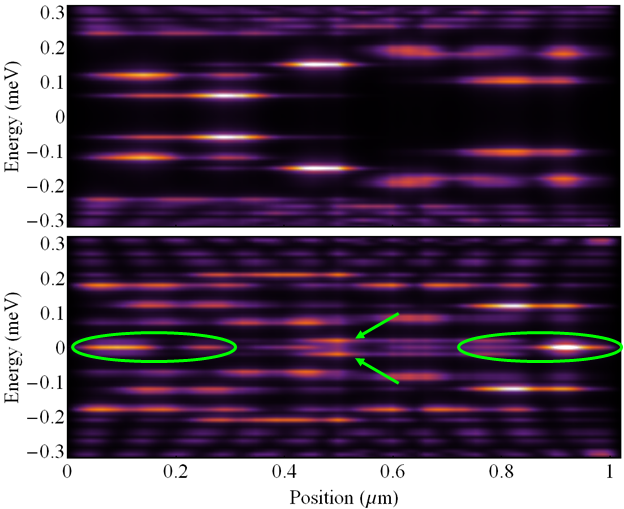}
\end{center}
\vspace{-3mm}
\caption{LDOS as a function of the position along the wire and energy for a system with chemical potential $\mu=0.5~$meV, effective SM-SC coupling $\widetilde{t}=50~$meV, and Zeman field $\Gamma=0.4~$meV (top) or $\Gamma=0.9~$meV (bottom). The highlighted features in the bottom panel are generated by the lowest energy modes shown in Fig. \ref{FIG17}. Notice that the pair of Majorana modes localized near the opposite ends of the wire hybridizes with the low-energy ABS localized near the center of the wire (arrows). By comparison with Fig. \ref{FIG16}, the features in the top panel ($\Gamma=0.4~$meV) tend to be more localized and asymmetric, which is an indication of stronger effective disorder.}
\label{FIG18}
\vspace{-1mm}
\end{figure}

To better understand the nature of the lowest energy modes, we calculate the LDOS for $\Gamma=0.4~$meV and $\Gamma=0.9~$meV (vertical cuts in Fig. \ref{FIG17}). The results are shown in Fig. \ref{FIG18}. For $\Gamma=0.4~$meV, the lowest energy modes are finite energy ABSs localized near the ends of the system (top panel in Fig. \ref{FIG18}). The lowest energy modes at  $\Gamma=0.9~$meV correspond to a pair of Majorana modes and a low-energy ABS (see bottom panel). This is similar to the situation discussed before in the context of Fig. \ref{FIG16}, but with the significant difference that the energy of the ABS is much lower. Also,  notice that the Majorana modes have finite spectral weight in the region where the ABS is localized, which indicates that local perturbations in this region may couple these modes. 
These features indicate that the effective (proximity-induced) disorder characterizing this system is stronger than that characterizing the system in Fig. \ref{FIG16}. A further indication of stronger effective disorder is that the features in the top panel of Fig. \ref{FIG18}  are more localized than the features in the top panel of Fig. \ref{FIG16}, although they correspond to the same value of $\Gamma$. Finally, note the strong asymmetry between the Majorana modes localized near the end of the wire, as compared to the more symmetric features in Fig. \ref{FIG16}. We emphasize that the low-energy ABS (marked by arrows in Fig. \ref{FIG18}) is localized away from the ends of the system and, consequently, remains practically ``invisible'' in (local or non-local) transport measurements. Increasing the length of the system will lead to a proliferation of this type of disorder-induced low-energy ABSs. Note that transport gaps visible in an end-to-end charge transport measurement \cite{Danon2020,Pan2021,Poschl2022} are associated with higher energy delocalized (bulk) states and do not characterize the quasiparticle gaps associated with the ``invisible'' low-energy ABSs. In other words, observing correlated zero-energy peaks in the differential conductance measured at the ends of the wire and a finite gap in the end-to-end transport does not eliminate the possibility of having ``invisible'' disorder-induced low-energy ABSs localized throughout the system.

Further increasing the effective SM-SC coupling results in features that are even more localized and more asymmetric  and generates low-energy ABSs that do not acquire a gap larger than the typical inter-state energy spacing. This indicates that increasing the effective SM-SC coupling (i.e., $\widetilde{t}$ or $\Lambda$) enhances the effective proximity-induced disorder \cite{Hui2015,Cole2016}. We emphasize that this enhancement has nothing to do with the physical disorder, which is the same (in this case the surface roughness of the SC film), but with changing the effective coupling $\widetilde{t}$, which can be done, for example, using a back gate potential that ``pushes'' the SM wave functions toward (or away from) the SM-SC interface. We also note that the maximum values of the SM-SC coupling consistent with weak induced disorder depend on the length scales characterising the physical disorder, as shown by a comparison between the results discussed in this section and the additional results presented in Appendix \ref{AppD}, which correspond to a different surface roughness profile. 

\section{Conclusion and discussion}\label{S5}

We have developed a nonperturbative theory of the proximity effect induced by thin SC films with strong (surface) disorder based on the exact numerical treatment of a 3D microscopic model. The challenge associated with the high numerical cost is addressed by combining a recursive Green's function method with a ``patching'' approach that exploits the quasi-local nature of the interface Green's function in the presence of (strong) disorder. 
Focusing on SC-SM hybrid structures, we find that the Fermi surface mismatch combined with the strong confinement-induced quantization of the transverse modes strongly suppress the proximity effect generated by clean thin films, by up to three orders of magnitude as compared to the proximity effect generated by a bulk SC, and make it strongly dependent on the film thickness $L_z$. A random onside model of surface disorder previously discussed in the literature produces a moderate enhancement of the proximity-induced superconductivity and does not suppress significantly the strong, non-monotonic thickness dependence.
However, we have constructed a model of surface roughness capable of efficiently mixing the ``clean'' SC modes, which results in a dramatic enhancement of the proximity effect (by up to three orders of magnitude), and a complete suppression of the strong dependence on  $L_z$ over significant ranges of thickness values. We find that, within these ranges, the proximity-induced superconductivity generated by a thin SC film is practically similar to the superconductivity induced by a bulk SC. These results hold as long as the surface roughness affects at least two atomic layers at the surface of the thin SC film, or at the interface of the superconducting metal with a surface oxide. Thus, we explicitly demonstrate that the presence of strong SC disorder is essential to inducing a robust proximity effect using thin SC films.

The other face of strong SC disorder is the emergence of induced effective disorder in the semiconductor. We investigate in detail the effect of the induced disorder on the emergent topological superconductivity and the associated low-energy modes. We demonstrate that the presence of SC surface roughness, which leads to a robust proximity effect, is consistent with the emergence of topological superconductivity and Majorana modes, provided the SC-SM coupling does not exceed values corresponding to an induced (zero field) gap up to about $0.6\Delta_0$, where $\Delta_0$ is the parent SC gap. Stronger values of the SC-SM coupling lead to the emergence of disorder-induced low-energy ABSs, which can be localized away from the ends of the system and remain ``invisible'' in a transport measurement. In addition, the presence of stronger induced disorder associated with a larger SC-SM coupling leads to the emergence of localized asymmetric spectral features. 

The presence of SC disorder both helps and hinders the realization of topological superconductivity in SM-SC hybrid Majorana platforms, with the competing effects being affected differently by various disorder parameters (e.g., characteristic length scales). The induced effective disorder also depends on the SM-SC coupling. To consistently realize structures with SC disorder capable 
to efficiently overcome the Fermi surface mismatch while inducing weak effective disorder in the SM nanowire, consistent with the emergence of a robust  topological SC gap, experimentalists must strike a careful balance through appropriate materials science development and characterization and careful device engineering. Based on our work, we cannot rule out the possibility  that the experimental finding of many SM-SC hybrid samples manifesting no SC proximity effect is actually arising from the Fermi surface mismatch problem not being fully compensated by the SC film disorder.  The subject merits careful materials science investigations, since, as we find theoretically, many details would matter.

We believe that our findings explain much of the existing phenomenology of SC-SM hybrid Majorana structures and clearly establish that, while the SM disorder must be eliminated as much as possible, SC disorder is not only beneficial, but in fact essential to the manifestation of robust proximity-induced superconductivity in the semiconductor and, consequently, to the realization of topological Majorana zero modes. However, this study also emphasizes the key importance of a detailed experimental characterization of the SC film. Outstanding issues include the characterization of SC disorder (e.g., the characteristic length scales), the systematic study of the thickness dependence of the proximity effect, the control of the effective SC-SM coupling, including the possibility of simultaneously controlling both the SC-SM coupling and the SM chemical potential, the detection of ``hidden'' disorder-induced low-energy ABSs, or the possibility of increasing the thickness of the SC system by engineering layered SC films involving two (or more) superconductors with optimized properties. Our work provides valuable theoretical tools for dealing with these issues, but solving them will ultimately require a substantial experimental effort. 
It is ironic that the very first paper dealing with SC disorder in SC-SM hybrid structures \cite{Potter2011} had the ominous (but incorrect \cite{Potter2011a,Lutchyn2012}) conclusion that  SC-SM structures are unsuitable for realizing topological superconductivity because of strong SC disorder and now, 11 years later, our work concludes that strong SC disorder not only is not detrimental, but, in fact, is essential to realizing topological superconductivity and Majorana zero modes in SC-SM structures. With a strong warning: details matter if one attempts to fabricate a functional quantum device; the details regarding the strong disorder in thin SC films remain an outstanding experimental challenge.

\section*{Acknowledgement}

This work is supported by the Laboratory for Physical Sciences and by NSF-2014156.

\appendix
\counterwithin{figure}{section}

\section{System with onsite random disorder} \label{AppA}

Here, we provide additional results for the relative amplitude $\widetilde{g}$ of a thin SC film with onsite random disorder at the surface. Focusing on the simplified 2D disorder model, we investigate the dependence on the disorder amplitude, the impact of changing the lattice constant and the thickness of the oxide layer, and the behavior of the system in the ultra-strong disorder limit.

\begin{figure}[tb]
\begin{center}
\includegraphics[width=0.48\textwidth]{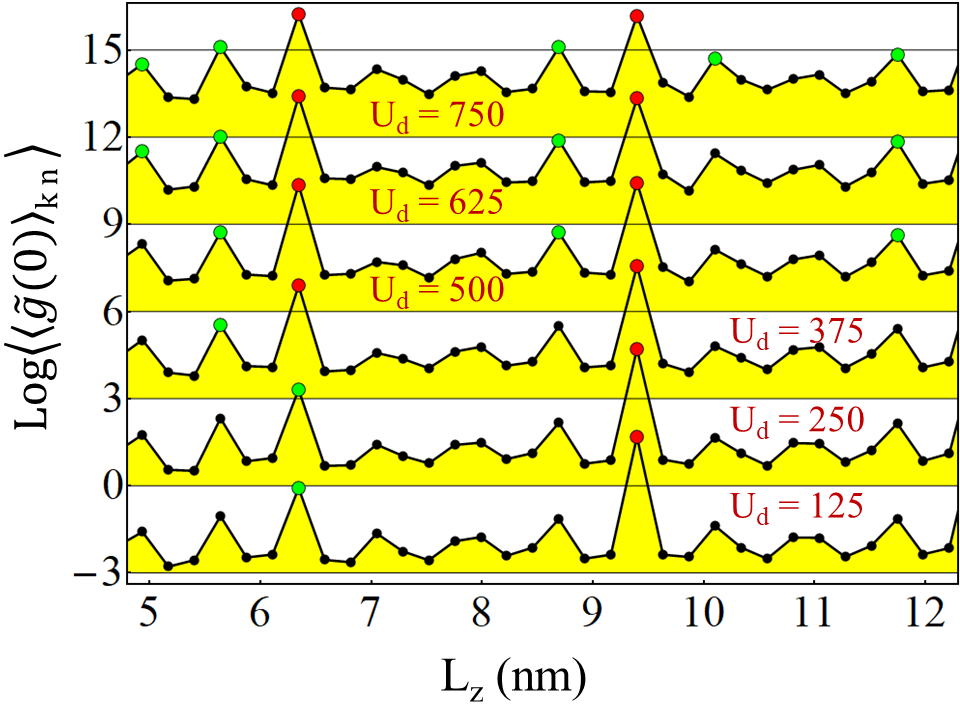}
\end{center}
\vspace{-3mm}
\caption{Dependence of the average Green's function amplitude $\widetilde{g}$ on the film thickness $L_z$ for different values of the amplitude $U_d$ of a 2D onsite random potential located within $2~$nm from the surface. The relative amplitude $\widetilde{g}$ is averaged over the wave vector $k$ (with $0\leq k \leq 0.25~$nm$^{-1}$),  the transverse mode $n_y$ (with $1\leq n_y \leq 5$), and 25 disorder realizations. The system has infinite length, width $L_y=50.2~$nm and is defined on a lattice with $a=0.38~$nm and $c=0.117$nm. The values of $U_d$ are given in meV and the corresponding curves are shifted (by multiples of 3), for clarity. The curves for $U_d=250~$meV and $U_d=500~$meV correspond to the results shown in Fig. \ref{FIG5}.}
\label{FIGA1}
\vspace{-1mm}
\end{figure}

First, we calculate the dependence of the average Green's function amplitude $\widetilde{g}$ on the film thickness for different values of the disorder amplitude $U_d$ for a system with 2D onsite random disorder within $2~$nm from the surface. The results shown in Fig.  \ref{FIGA1} clearly indicate that stronger disorder generates more enhancement of the interface Green's function amplitude, hence a stronger superconducting proximity effect. Nonetheless, the presence of disorder does not eliminate the strong dependence on the film thickness. We emphasize that the results in Fig.  \ref{FIGA1} (and other similar figures) show the logarithm (base 10) of $\widetilde{g}$, so a variation by 1 in the figure implies a variation by one order of magnitude of $\widetilde{g}$. Also note that the values of the lattice constants used in the calculation are the same as in Fig. \ref{FIG5}, which implies that $K_d \lesssim 1000~{\rm meV}^2{\rm nm}^3$, similar to the values used in Refs. \cite{Antipov2018,Winkler2019}, corresponds to $U_d\lesssim 250~$meV. The strong thickness dependence persists up to significantly larger disorder potential amplitudes (e.g., $U_d=750~$meV in Fig. \ref{FIGA1}), despite the substantial disorder-induced enhancement (by up to two orders of magnitude) of the interface Green's function amplitude.

\begin{figure}[t]
\begin{center}
\includegraphics[width=0.48\textwidth]{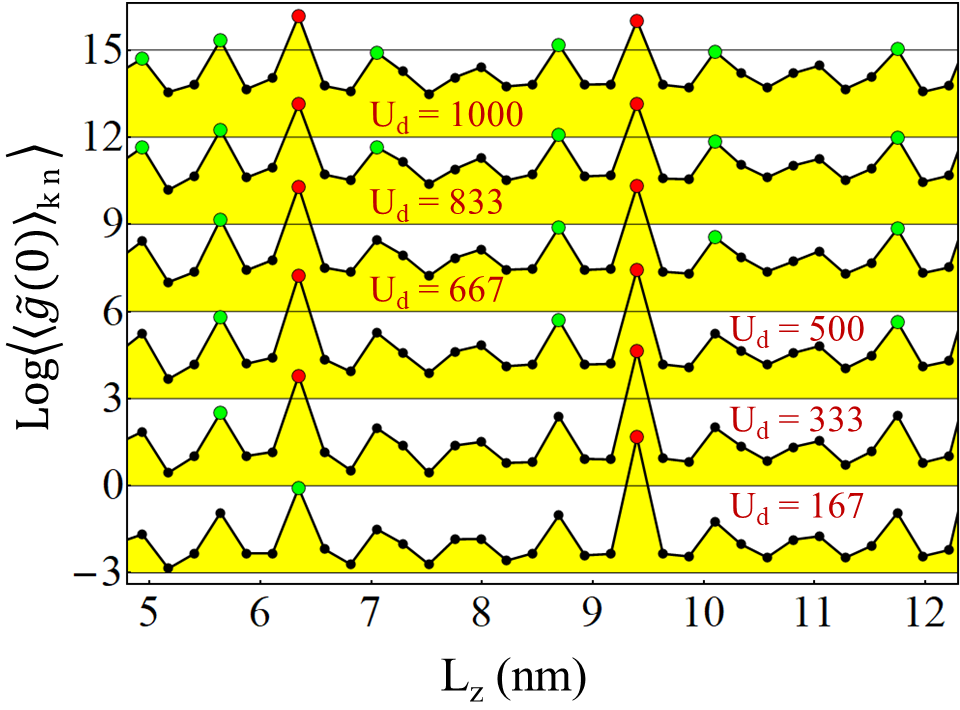}
\end{center}
\vspace{-3mm}
\caption{Same as in Fig. \ref{FIGA1}, but for a system with lattice constants $a=0.285~$nm and $c=0.117$nm. The values $U_d$ of the random potential amplitude correspond to the same value of $K_d$ as the corresponding curves in Fig. \ref{FIGA1}. For example $U_d=333~$meV  corresponds to $K_d\approx 1060~{\rm meV}^2{\rm nm}^3$ (same as $U_d=250~$meV in Fig. \ref{FIGA1}), while $U_d=1000~$meV  corresponds to $K_d\approx 9540~{\rm meV}^2{\rm nm}^3$ (same as $U_d=750~$meV in Fig. \ref{FIGA1}).}
\label{FIGA2}
\vspace{-1mm}
\end{figure}

Next, we investigate the impact of changing the (in-plane) lattice constant $a$ on the disorder-induced effects. Fig.  \ref{FIGA2} shows the dependence of the average relative amplitude  $\widetilde{g}$ on the film thickness $L_z$ and on the strength of the disorder potential for a model characterized by a lattice constant smaller than the one in Fig. \ref{FIGA1}  ($a=0.285~$nm, instead of $a=0.38~$nm). The reduction of the lattice constant (by a factor of $3/4$) is ``compensated'' by an enhancement of the random potential amplitude  (by a factor of $4/3$), to obtain an effective disorder strength equivalent (i.e., having the same value of $K_d$) to the disorder strength of the corresponding curves in Fig. \ref{FIGA1}. Comparison with Fig. \ref{FIGA1} (and the semi-quantitative agreement between the corresponding curves) shows explicitly that models characterized by the same value of $K_d = a^2 c ~\!U_d^2$ correspond to systems with ``equivalent'' effective disorder. 

At this point it is worth noting an important technical aspect. The disorder-induced enhancement of the interface Green's function is essentially due to a disorder-induced coupling of the long wavelength in-plane modes $(n_x, n_y)$ to modes near the Fermi level having arbitrary quantum numbers. This requires the presence of quantum states (with arbitrary quantum numbers) near the Fermi energy. However, if the the energy of the transverse mode immediately below the Fermi level is $\epsilon_{n_z} < \epsilon_F -8t$, where $8t$ is the bandwidth associated with the in-plane modes, there are no low-energy modes available. In turn, the bandwidth depends on the in-plane lattice constant as $8t=4\hbar^2/m_e a^2$. Consequently, when using  the onsite random disorder model one should choose a value of the lattice constant $a$ that is small enough to ensure the presence of low-energy states (near the Fermi level) for all $L_z$ values of interest. Further reducing the lattice constant will generate ''equivalent'' disorder potentials for any given value of $K_d$, as illustrated above.

\begin{figure}[t]
\begin{center}
\includegraphics[width=0.48\textwidth]{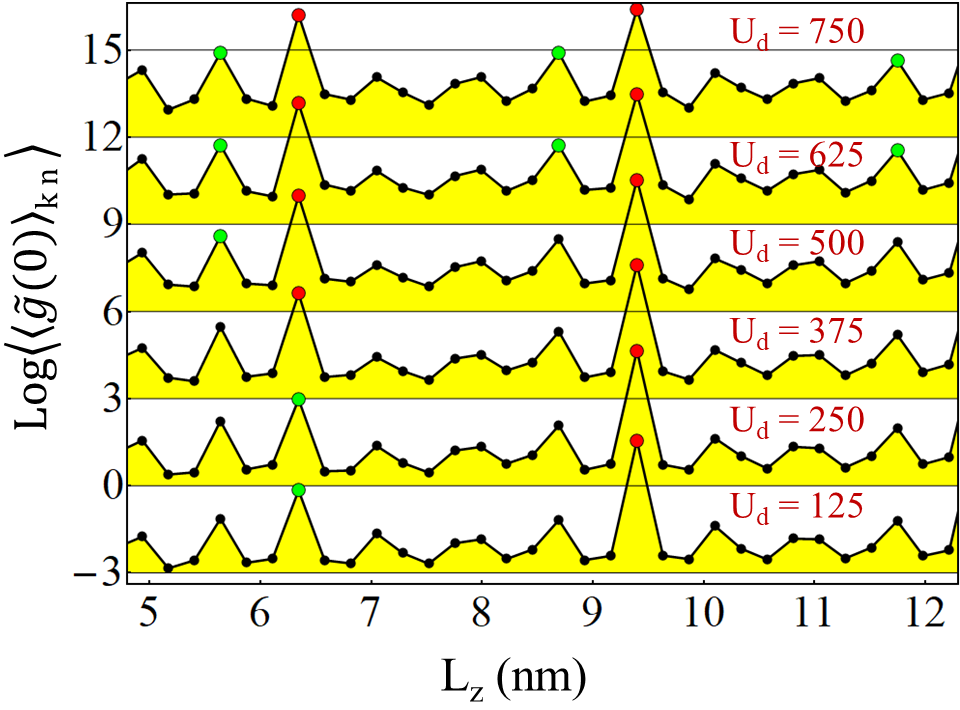}
\end{center}
\vspace{-3mm}
\caption{Same as in Fig. \ref{FIGA1}, but for a system with onsite random disorder within $1~$nm from the surface. Reducing the thickness of the oxide layer (by a factor of two) slightly reduces the disorder-induced enhancement of $\widetilde{g}$, but does not affect its thickness dependence.}
\label{FIGA3}
\vspace{-1mm}
\end{figure}

The effect of reducing the thickness of the oxide layer by a factor of two (as compared to the system in Fig. \ref{FIGA1}) is illustrated in Fig.  \ref{FIGA3}. While direct comparison with Fig. \ref{FIGA1} shows a slight reduction of the disorder-induced enhancement of the relative amplitude, the  thickness dependence of $\widetilde{g}$ is basically not affected. In particular, for a given value of $K_d$, the order of magnitude of the (average) relative amplitude is determined by the total film thickness, $L_z$, which includes the oxide layer. One can consistently identify the values of $L_z$ associated with large (or small) relative amplitudes in Figs. \ref{FIGA1}-\ref{FIGA3}, which correspond to systems with 2D random potential disorder, or in Fig. \ref{FIG7} (system with 3D random potential disorder) and one can ultimately trace them back to the clean case shown in Fig. \ref{FIG3}. This demonstrates that, for reasonable values of the disorder strength, the random potential model cannot eliminate the strong dependence of the interface Green's function (hence, the dependence of the proximity-induced effect) on the thickness of the SC film.

\begin{figure}[t]
\begin{center}
\includegraphics[width=0.48\textwidth]{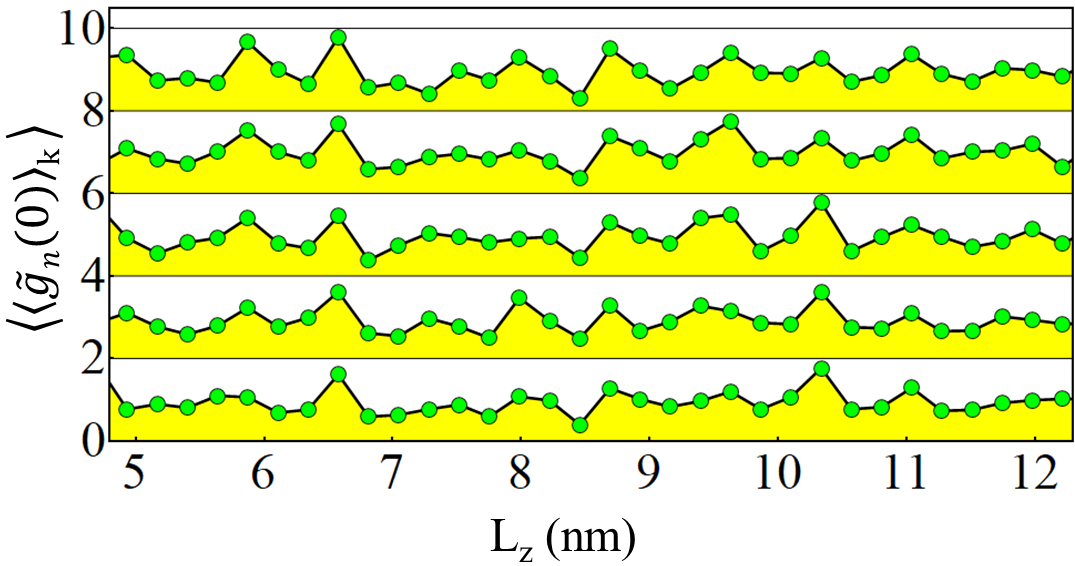}
\end{center}
\vspace{-3mm}
\caption{Thickness dependence of the diagonal relative amplitude $\widetilde{g}_{n_y}(\omega, k)$ averaged over $k\in[0, 0.25~{\rm nm}^{-1}]$ and over 50 different disorder realizations for an infinitely long  SC system with $L_y=50.2~$nm and strong 2D random potential disorder of amplitude $U_d=10~$eV within $2~$nm from the surface.  The curves corresponding to the lowest five $n_y$ modes are shifted by $2(n_y-1)$, for clarity. Note that we represent the average relative amplitude, rather than the logarithm of this quantity. The system is in the ``optimal regime'' for all values of $L_z$.}
\label{FIGA4}
\vspace{-1mm}
\end{figure}

Finally, in Fig.  \ref{FIGA4} we show that, within this model, the problem regarding the strong thickness dependence of the interface Green's function 
can only be eliminated if we consider an extremely strong disorder potential, e.g., $U_d=10~$eV, which is rather unphysical. In this limit, the dependence on $L_z$ is weak and the disordered thin film is practically equivalent to a bulk superconductor, i.e., $\langle\langle\widetilde{g}_{n_y}\rangle_k\rangle \sim 1$.
Note that, in addition to enhancing the low-$\widetilde{g}$ values corresponding to the clean system (black points in Fig. \ref{FIG3}) by $1-3$ orders of magnitude, ultra-strong disorder also reduces the large $\widetilde{g}$ corresponding to $L_z=9.4~$nm (red points in Fig. \ref{FIG3}), so that the relative amplitude is in the ``optimal regime'' independent of the film thickness.

\section{Modeling the surface roughness} \label{AppB}

\begin{figure}[t]
\begin{center}
\includegraphics[width=0.48\textwidth]{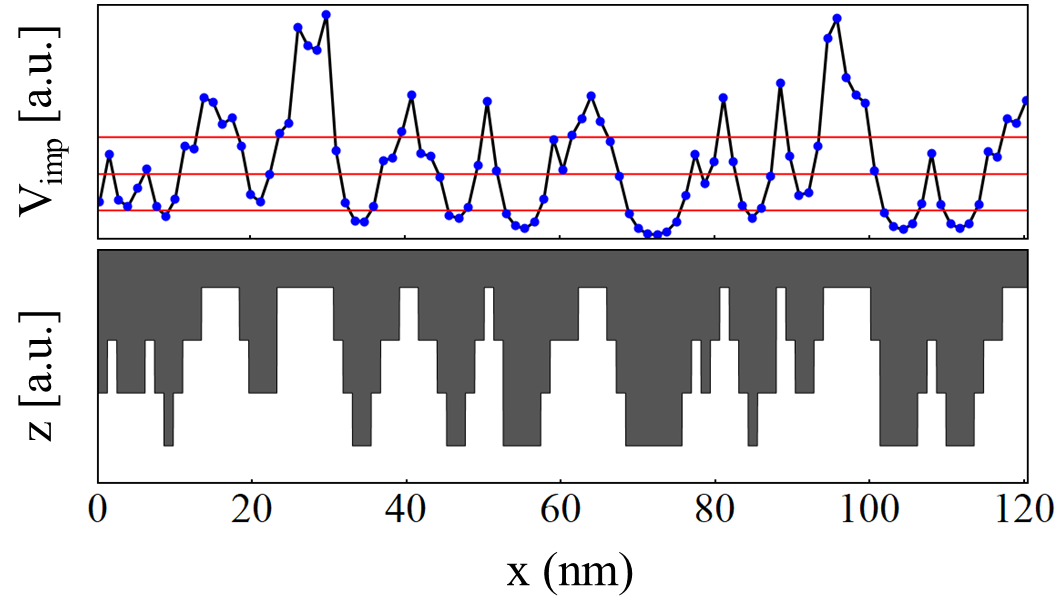}
\end{center}
\vspace{-3mm}
\caption{Schematic representation of the surface roughness modeling for a 1D ``surface''. {\em Top}: The ``impurity potential'' generated by Eq. (\ref{EqB1}) is separated by the red lines into four regions, each containing approximately 25$\%$ of the lattice sites. {\em Bottom}: The shaded region ``above'' the random potential is non-superconducting (e.g., insulator or vacuum), while the white region represents the rough surface of the superconductor. Each step (in the $z$ direction) corresponds to one atomic layer; the surface roughness in this schematic example affects the top three layers.}
\label{FIGB1}
\vspace{-1mm}
\end{figure}

Let us assume that surface roughness affects the top $r$ atomic layers of the SC film. To model the rough surface, we first generate a fictitious 2D ``impurity potential'' defined as 
\begin{equation}
V_{imp}({\bm i}) = \sum_{p=1}^{N_{imp}} \exp\left(-\frac{|{\bm i}-{\bm i}_p|}{\lambda}\right),  \label{EqB1}
\end{equation}
where ${\bm i} = (i_x, i_y)$ labels lattice sites in the $x-y$ plane, ${\bm i}_p$ describes the (randomly generated) position of ``impurity'' $p$, $N_{imp}$ is the the total number of ``impurities'', and $\lambda$ is a parameter that controls the characteristic length scale of the potential. Next, we ``cut'' the ``impurity potential'' by $r$ planes at heights determined by the percentage of superconducting phase corresponding to each layer, as shown schematically in Fig. \ref{FIGB1}. Finally, for all points that are ``above'' the ``impurity potential'' (see Fig. \ref{FIGB1}) we add a local potential $V_{dis}(i_x, i_y, i_z) = \epsilon_F + U$, with $U=4~$eV. Basically,  $V_{dis}$ confines the electrons into the superconducting region, i.e., it pushes them out of the non-superconducting region (e.g., the black region in   Fig. \ref{FIGB1}). We note that the number of atomic layers affected by surface roughness, hence the amplitude $\delta L_z$ of the thickness fluctuations, is determined by the parameter $r$ (i.e., the number of ``cutting'' planes in our construction), while the in-plane characteristic length scale is controlled by the ``impurity'' density, $n_{imp}=N_{imp}/L_x L_y$, and by the parameter $\lambda$ in Eq. (\ref{EqB1}). 

\begin{figure}[t]
\begin{center}
\includegraphics[width=0.48\textwidth]{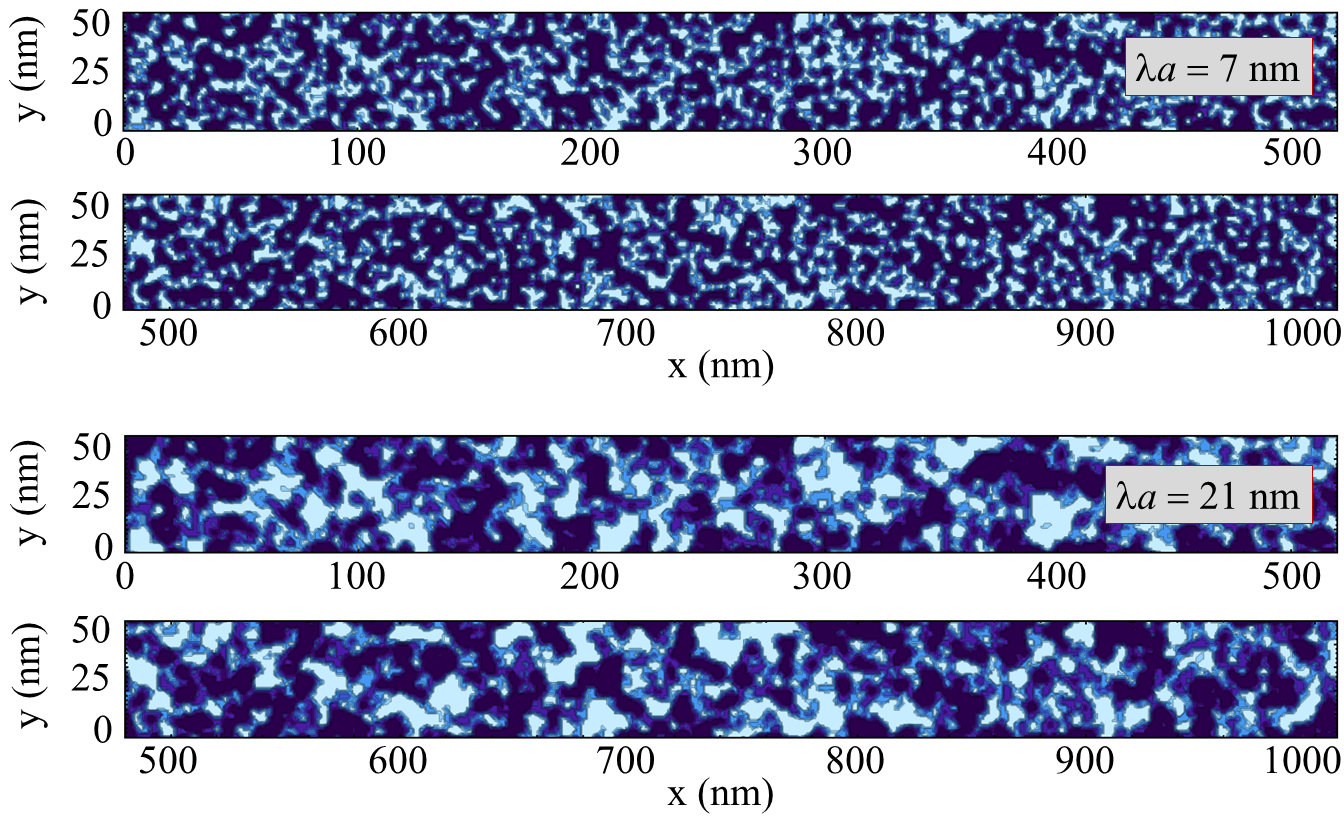}
\end{center}
\vspace{-3mm}
\caption{Surface roughness realizations corresponding to a specific ``impurity'' distribution and two values of the parameter $\lambda$ [see Eq. (\ref{EqB1})].  Note that  the wire of length  $L_x=1.025~\mu$m  is broken into two segments, for clarity. Also note that the top panel is the same as Fig. \ref{FIG1}(c). Roughness affects the top three atomic layers (i.e., $r=3$), which are only partially in the superconducting phase.  Four different shades of blue correspond to different values of the {\em superconducting} wire thickness (i.e., the local thickness of the superconducting region) each covering approximately $25\%$ of the surface. The difference between the maximum (light blue) and minimum (dark blue) values is approximately $0.7~$nm (i.e., three atomic layers).}
\label{FIGB2}
\vspace{-1mm}
\end{figure}

To illustrate the role of the parameter $\lambda$, we consider a SC wire of length $L_x=1.025~\mu$m and width $L_y=51.3~$nm and a specific random distribution of ``impurities'', $\{{\bm i}_1, \dots, {\bm i}_{N_{imp}}\}$, corresponding to a density $n_{imp} = 0.058~{\rm nm}^{-2}$. We take $r=3$, i.e., the top three atomic layers are only partially in the superconducting phase. The surface roughness profiles corresponding to $\lambda a = 7~$nm and  $\lambda a = 21~$nm are shown in Fig. \ref{FIGB2} top and bottom panels, respectively. Clearly, increasing $\lambda$ generates surface roughness with larger in-plane characteristic length. Repeating the procedure for a different  random distribution of ``impurities'' (all other parameters being the same) allows us to generate random surface roughness profiles with the same general characteristics, in particular same $\delta L_z$ and in-plane characteristic length. We note that the surface roughness realization shown in the top panel of Fig. \ref{FIGB2} is used in the numerical calculations presented in Sec. \ref{S3B2} and Sec. \ref{S4}. In addition, a surface roughness profile similar to that shown in the bottom panel of Fig. \ref{FIGB2}, but having $r=2$ (i.e., affecting the top two atomic layers) is used in the calculations presented in Appendix \ref{AppD}.

\section{The patching approach} \label{AppC}

To efficiently simulate large systems, i.e., finite 3D SC films with large interface areas, we exploit the fact that in the presence of disorder the interface Green's function is short ranged and its values in the vicinity of a given point do not depend on the properties of the system (e.g., on the disorder potential values) at distances much larger than the characteristic decay length of the Green's function, which is on the order of a few nanometers. Consequently,  
we can calculate the (quasi-local) Green's function by explicitly simulating a sufficiently large ``patch'' around the vicinity of interest. 

\begin{figure}[t]
\begin{center}
\includegraphics[width=0.48\textwidth]{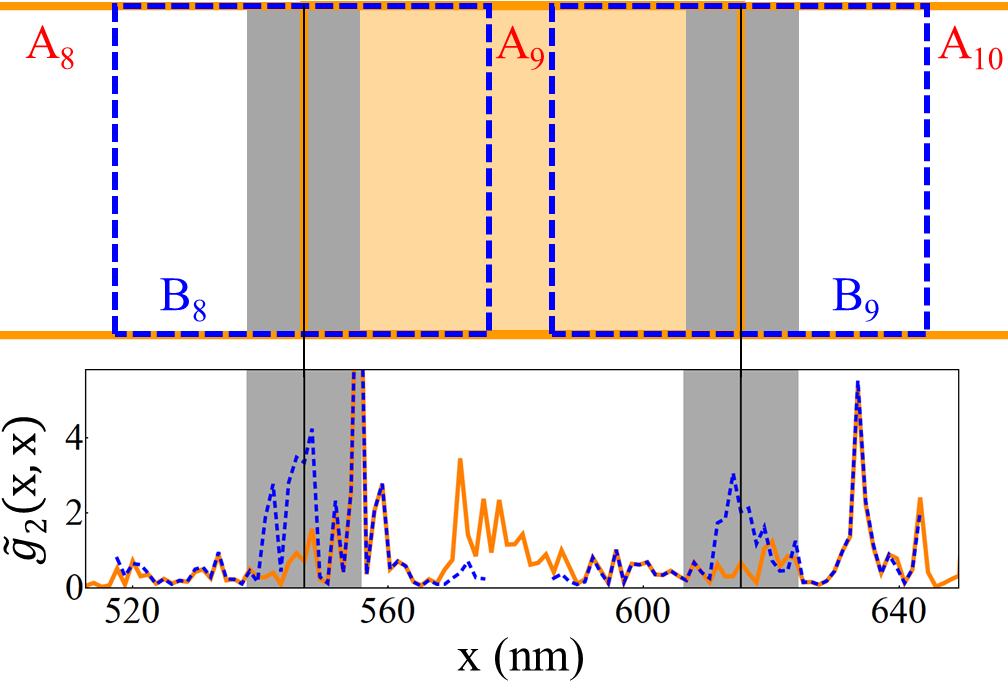}
\end{center}
\vspace{-3mm}
\caption{{\em Top}: Schematic representation of patch $A_9$ and its neighbors, $A_8$ and $A_{10}$ (orange lines) and of auxiliary patches $B_8$ and $B_9$ (dashed blue lines) that cover the boundary regions (gray areas) of the A-patching. The dimensions of the A-patches are  $\ell_x \times  \ell_y = 68.4 \times 51.3~$nm, while B-patches have dimensions $\ell_x \times  \ell_y = 58.6 \times 51.3~$nm. Note that $\ell_y = L_y$, i.e., each patch covers the entire width of the wire. The light orange shading corresponds to the ``bulk'' of patch $A_9$. {\em Bottom}: Dependence of the relative amplitude $\widetilde{g}_{n_y}(\omega; x,x)$ on the longitudinal position $x$ along the wire for a film of thickness $L_z=8~$nm. The values calculated using the A-patches are represented by the orange line, while the blue dashed lines are obtained using the B-patches. The final values of $\widetilde{g}$ (see the middle panel of Fig. \ref{FIG11}) within the ``bulk'' regions are obtained using the A-patching (orange lines), while within the boundary regions (gray areas)  we use the B-patching results (blue dashed lines). Note that in the regions that are sufficiently far away from the boundaries of both types of patches the two calculations give the same result (i.e., the orange and blue dashed lines are on top of each other).}
\label{FIGC1}
\vspace{-1mm}
\end{figure}

For specificity, let us consider the finite 3D SC film shown in Fig. \ref{FIG1}, which has $L_x=1.025~\mu$m, $L_y = 51.3~$nm and a surface roughness as shown in  Fig. \ref{FIG1}(c). First, we completely cover the wire with 15 patches ($A_1, A_2, \dots, A_{15}$) of dimensions $\ell_x \times  \ell_y = 68.4 \times 51.3~$nm and calculate the interface Green's function within each patch. In the top panel of Fig. \ref{FIGC1} we represent schematically patch $A_9$ and its neighbors (orange lines). Given the quasi-locality of the Green's function, the calculated values are accurate within the ``bulk'' of each patch (the orange shaded area in Fig. \ref{FIGC1} for path $A_9$), but deviations are expected within the boundary areas (gray shaded region in Fig. \ref{FIGC1}). To obtain the correct values of $\widetilde{G}$ in the boundary regions, we use 14 auxiliary patches, $B_1, B_2, \dots, B_{14}$, that cover these regions and calculate $\widetilde{G}$ within each auxiliary patch. A specific example is shown in the bottom panel of Fig. \ref{FIGC1}. We emphasize that for points $x$ that are far-enough from the boundaries of both types of patches the results obtained using A-patches coincide with those obtained using B-patches. We explicitly checked that this property holds for all relevant non-local Green's functions  $\widetilde{G}_{n_y}(\omega; x, x^\prime)$, i.e., for $\delta x = |x-x^\prime| \lesssim 16~$nm (see Fig. \ref{FIG10}). Finally, we note that our patching approach can be generalized to thin films with 2D (rather than quasi-1D) surfaces/interfaces of arbitrary size and shape. 

\section{Superconducting thin film with surface roughness: additional results} \label{AppD}

\begin{figure}[t]
\begin{center}
\includegraphics[width=0.48\textwidth]{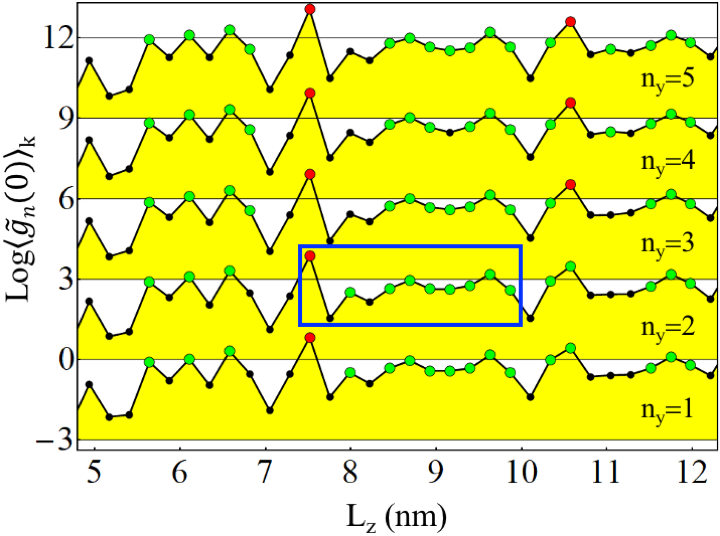}
\end{center}
\vspace{-3mm}
\caption{ Dependence of the averaged relative amplitude, $\langle\widetilde{g}_{n_y}(0)\rangle_{k_x}$, on the SC film thickness, $L_z$, for a  system with surface roughness similar to the specific realization shown in the lower panel of Fig. \ref{FIGB2}, but having $r=2$ (i.e., $\delta L_x= \pm 0.23~$nm). The wire has length $L_x=1.025~\mu$m and width $L_y=51.3~$nm.  The relative amplitude for the lowest five $n_y$ modes  was averaged over $k_x$, with $0 \leq k_x \lesssim 0.25~$nm$^{-1}$, which corresponds to $n_x\leq 82$. The corresponding curves are shifted by $3(n_y-1)$, for clarity. The thickness dependence is much weaker than in the clean case (see Fig. \ref{FIG3}), but stronger than in Fig. \ref{FIG8}. The average amplitude  $\langle\widetilde{g}_{n_y}(0)\rangle_{k_x}$ is in the optimal regime for about 50$\%$ of the $L_z$ values. The corresponding disorder-induced enhancement is dramatic, by up to three orders of magnitude. The proximity effect generated by systems corresponding to the parameters inside the blue rectangle is discussed in the context of Figs. \ref{FIGD3}-\ref{FIGD6}.}
\label{FIGD1}
\vspace{-1mm}
\end{figure}

To strengthen our conclusions regarding the effects of surface roughness, we consider a surface roughness profile obtained using the same random distribution of fictitious impurities as in the main text (see Appendix \ref{AppB} for details), but having a larger in-plane characteristic length scale (corresponding to $\lambda~\!a = 21~$nm, instead of $\lambda~\!a = 7~$nm) and a smaller transverse length scale ($\delta L_x= \pm 0.23~$nm corresponding to $r=2$, instead of $\delta L_x= \pm 0.35~$nm, which corresponds to $r=3$). The roughness profile is almost identical to the profile shown in the bottom panel of Fig. \ref{FIGB2}, except that it involves the top two atomic layers (instead of the top three atomic layers in Fig. \ref{FIGB2}). To evaluate the impact of surface disorder on the interface Green's function, we calculate the thickness dependence of the zero-frequency relative amplitude averaged over $k_x$ (with $0\leq k_x\lesssim 0.25~$nm$^{-1}$), similar to the calculation leading to the results shown in Fig. \ref{FIG8}. Comparing the results shown in Fig. \ref{FIGD1} with the results discussed in the main text we note that the roughness profile considered here is generating a dramatic disorder-induced enhancement of the interface Green's function (by up to three orders of magnitude), but it is less effective in suppressing the thickness dependence of $\widetilde{g}$ (as compared with the case discussed in the main text). Perhaps surprisingly, this is mostly due to the larger in-plane characteristic length scale, while reducing $\delta L_z$ (i.e., $r$) has a weaker impact. Without performing a systematic study of the dependence on the parameters $\lambda$ (which controls the in-plane length scale) and $r$ (which controls the transverse length scale) -- such a study involves an enormous numerical cost and is marginally relevant in the absence of detailed experimental information -- we have verified that (i) surface roughness is effective in generating a huge disorder-induced enhancement of the interface Green's function as long as $r\geq 2$ (i.e., if it affects at least two atomic layers) and (ii) the dependence of the disorder effect on $\lambda$ is non-monotonic (with a certain finite in-plane length scale generating the maximum effect). 

\begin{figure}[t]
\begin{center}
\includegraphics[width=0.48\textwidth]{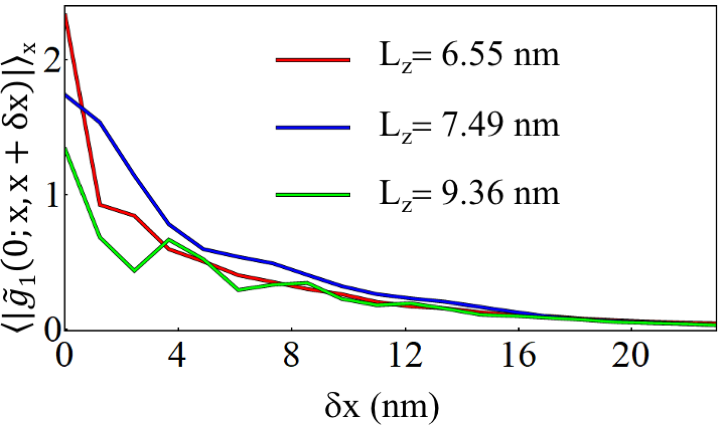}
\end{center}
\vspace{-3mm}
\caption{Dependence of the off-diagonal relative amplitude $|\widetilde{g}_{n_y}(\omega; x, x^\prime)|$ on the longitudinal distance $\delta x = |x^\prime - x|$ between two points. The quantity was averaged over all positions $x$ along the wire.
 Note the sharp decay of  $\widetilde{g}$ with increasing $\delta x$, which reveals that in the presence of surface disorder the interface Green's function becomes short ranged (i.e., quasi-local). Also note that the characteristic decay length is larger than the decay length characterising the system with surface roughness shown in Fig. \ref{FIG10}. 
 We have $\omega=0$, $n_y=1$, and three different film thicknesses, all other parameters being the same as in Fig. \ref{FIGD1}.}
\label{FIGD2}
\vspace{-1mm}
\end{figure}

Having established that the surface roughness profile studied here corresponds to a weaker ``effective surface disorder'' as compared to the case discussed in the main text, we investigate the real-space properties of the interface Green's function focusing on its non-local properties. Fig. \ref{FIGD2} shows the dependence of the off-diagonal elements  $\widetilde{g}_{n_y}(\omega; i_x, j_x)$ on the longitudinal distance $\delta x =|i_x-j_x|a$ for the mode $n_y =1$ and three different values of the film thickness. Note that the decay of the relative amplitude with increasing $\delta x$ has a characteristic length scale larger than the characteristic length scale in Fig. \ref{FIG10}. We associate this larger length scale with the weaker effective disorder corresponding to this surface roughness profile. In other words, stronger effective disorder (generated by surface roughness) results not only in a weaker thickness dependence of the interface Green's function (which approaches the value corresponding to a bulk SC), but also in a more localized $\widetilde{G}$. Physically, strong disorder is very effective in scrambling the $k$ modes of the clean system and, consequently, the interface ``looses'' information about the spatial location of the surface and the interface Green's function of the thin SC film becomes similar to interface Green's function of a bulk SC.

\begin{figure}[t]
\begin{center}
\includegraphics[width=0.48\textwidth]{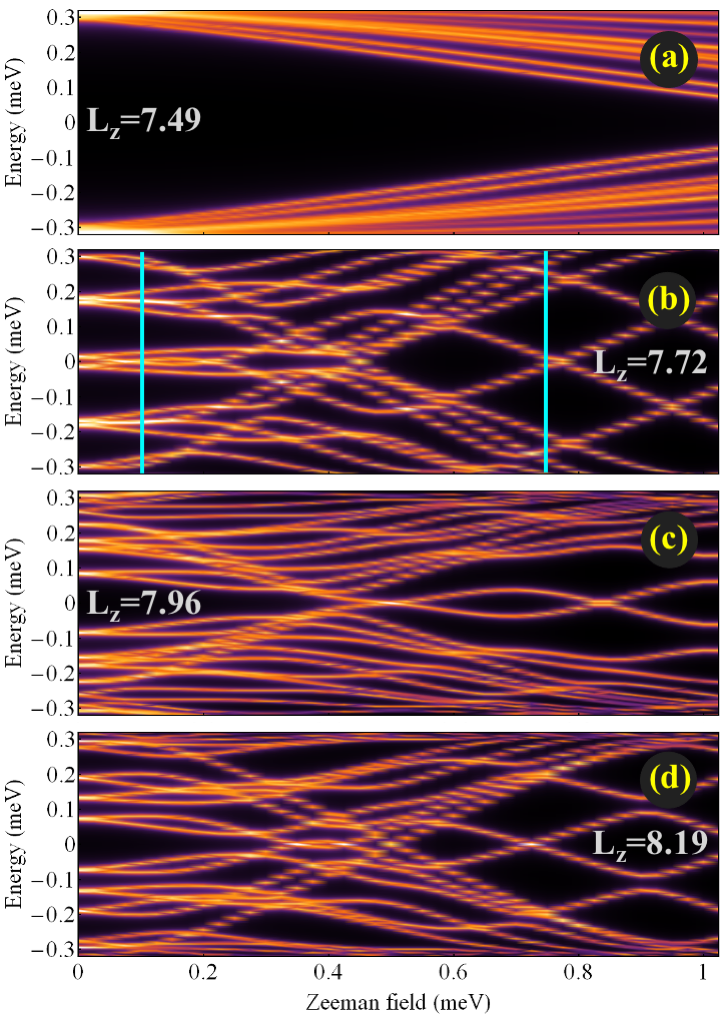}
\end{center}
\vspace{-3mm}
\caption{DOS as a function of Zeeman field and energy for a system with chemical potential $\mu=0.5~$meV and  effective SM-SC coupling  $\widetilde{t}=53.8~$meV ($\Lambda=0.4~$meV) for a sequence of four film thicknesses. The other system parameters are the same as in Fig. \ref{FIGD1}. The spectrum in panel (a) corresponds to a strongly coupled SM-SC system and has no low-energy states within the range of Zeeman fields investigated here. Panels (b) and (d) correspond to weakly coupled systems, while panel (c) shows the spectrum of a system with intermediate coupling. 
The LDOS corresponding to $\Gamma=0.1~$meV and $\Gamma=0.75~$meV -- vertical cuts in panel (B) -- is shown in Fig. \ref{FIGD4}.}
\label{FIGD3}
\vspace{-1mm}
\end{figure}

Next, we consider a hybrid SM-SC system and the effects of proximity-induced disorder on the induced superconductivity and the low-energy states, similar to the analysis in Sec. \ref{S4}. In Sec. \ref{S4} we have shown that the induced disorder strength is proportional to the ``total'' SM-SC coupling $\Lambda \langle\widetilde{g}(0)\rangle_{k_x}$. Here, we fix the SM-SC coupling, $\widetilde{t} = 53.8~$meV ($\Lambda=0.4~$meV), and focus on the thickness dependence of the proximity effect for the sequence of film thicknesses marked by the blue rectangle in Fig. \ref{FIGD1}.  We note that this value of the SC-SM coupling places the systems corresponding to green dots in Fig. \ref{FIGD1} in the intermediate coupling regime (i.e., total SM-SC coupling comparable to $\Delta_0$), the systems corresponding to black dots in the weak coupling regime, and the system with $L_z=7.49~$ (red dots in Fig. \ref{FIGD1}) in the strong coupling regime. 

\begin{figure}[t]
\begin{center}
\includegraphics[width=0.48\textwidth]{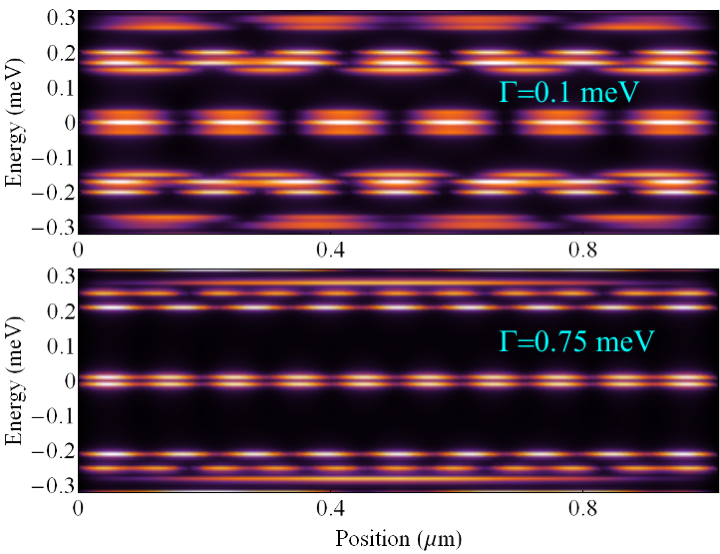}
\end{center}
\vspace{-3mm}
\caption{LDOS as a function of the position along the wire and energy for a system with weak SM-SC coupling having $L_x=7.72~$nm, chemical potential $\mu=0.5~$meV, effective SM-SC coupling $\widetilde{t}=53.8~$meV, and Zeman field $\Gamma=0.1~$meV (top) or $\Gamma=0.75~$meV (bottom), corresponding to the vertical cuts in Fig. \ref{FIGD3}(b). The other system parameters are the same  as in Fig. \ref{FIGD1}. Note that, as a result of negligible induced disorder, all states are delocalized and highly symmetric.}
\label{FIGD4}
\vspace{-1mm}
\end{figure}

\begin{figure}[t]
\begin{center}
\includegraphics[width=0.39\textwidth]{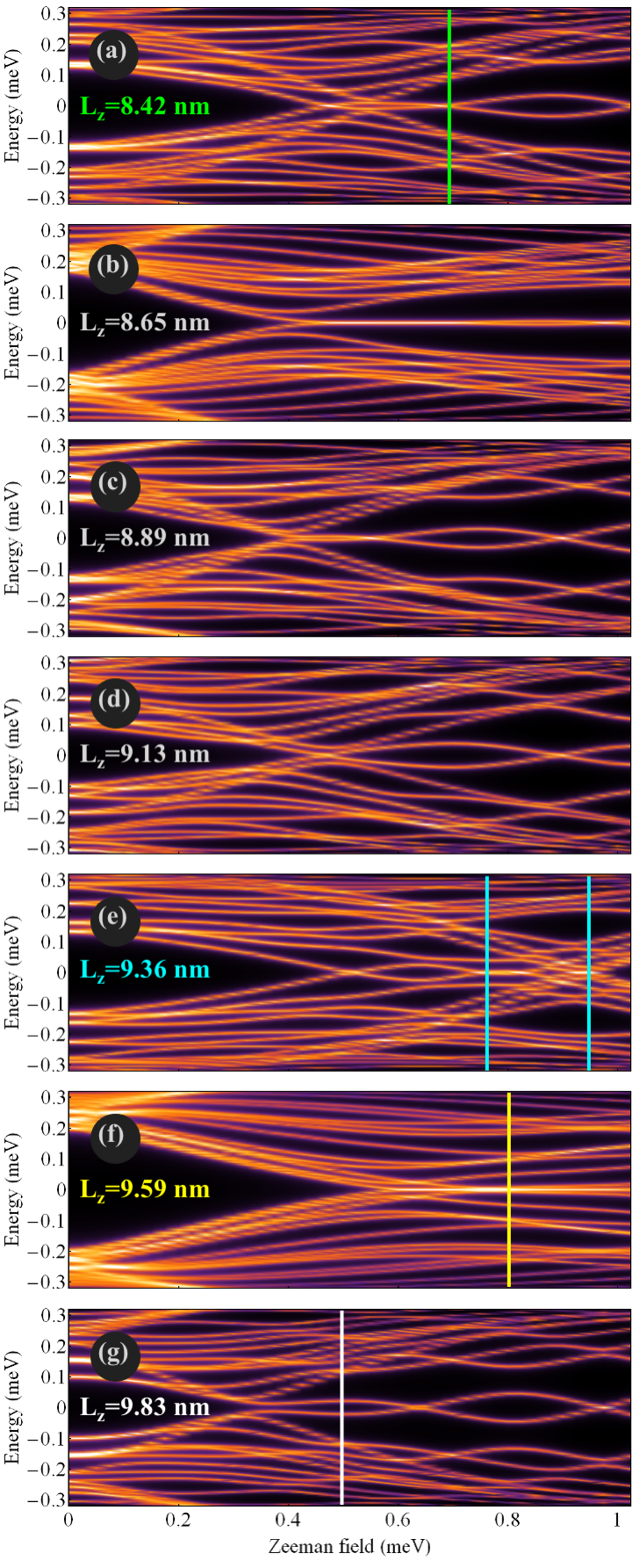}
\end{center}
\vspace{-3mm}
\caption{DOS as a function of Zeeman field and energy for a system with chemical potential $\mu=0.5~$meV, effective SM-SC coupling  $\widetilde{t}=53.8~$meV, and a sequence of seven film thickness values (also see Fig. \ref{FIGD1}). All systems in this sequence are in the intermediate coupling regime.}
\label{FIGD5}
\vspace{-1mm}
\end{figure}

We fix the chemical potential at $\mu=0.5~$meV and determine the dependence of the DOS on the Zeeman field $\Gamma$ for four thickness values: $L_z=7.49; 7.72; 7.96; 8.19~$nm (leftmost four points inside the blue rectangle in Fig. \ref{FIGD1}). The results shown in Fig. \ref{FIGD3} clearly reveal the fundamental problem of inducing (robust) superconductivity using thin SC films, particularly the large variations in the proximity-induced effect associated with variations of the film thickness by one atomic layer. If, for example, the SC film has a (maximum) thickness $L_z=7.49~$nm within the left side of the hybrid system and $L_z=7.72~$nm within the right side (with similar surface roughness throughout), the resulting hybrid system will be highly inhomogeneous, with no low-energy states revealed by any local measurement at the left end of the wire (e.g., tunneling conductance measurements) and no significant zero-field gap associated with local measurements at the right end of the wire [see Fig. \ref{FIGD3}(a-b)]. Obviously, such an inhomogeneous system cannot support Majorana zero modes at low-enough values of the Zeeman field, i.e., before the collapse of the SC gap. In addition, we note that the system with intermediate SM-SC coupling [panel (c)] is characterized by a low-energy mode with (relatively large) energy splitting oscillations for $\Gamma \gtrsim 0.45~$meV. As we will show below, this type of feature is generated by a pair of partially overlapping Majorana modes localized at the two ends of the wire.

\begin{figure}[t]
\begin{center}
\includegraphics[width=0.48\textwidth]{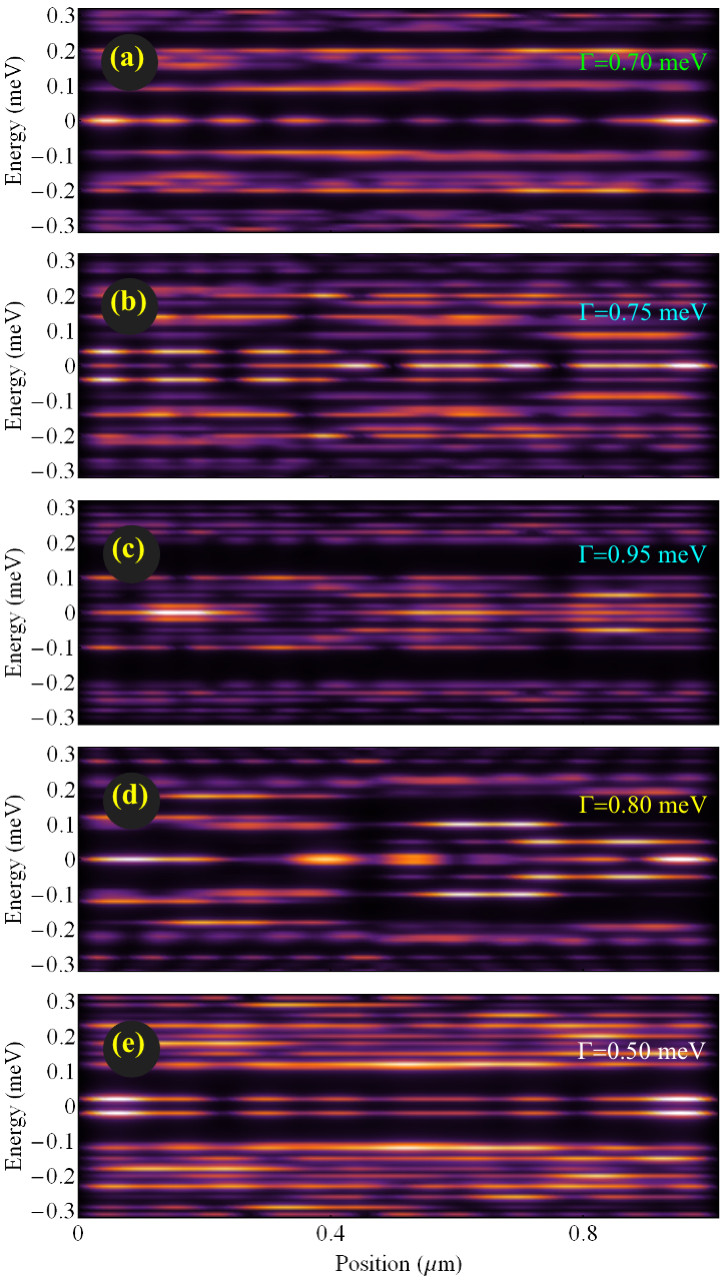}
\end{center}
\vspace{-3mm}
\caption{LDOS as a function of the position along the wire and energy for a system with chemical potential $\mu=0.5~$meV, effective SM-SC coupling $\widetilde{t}=53.8~$meV, and Zeman field and film thickness values corresponding to the vertical cuts in Fig. \ref{FIGD5}. Panels (a) and (e) show the presence of (partially overlapping) Majorana modes emerging inside a finite quasiparticle gap in a finite system with weak effective disorder. Panel (d) shows a pair of Majorana modes and a nearly-degenerate disorder- induced ABS localized near the middle of the wire. A proliferation of disorder-induced low-energy states is illustrated in panels (b) and (c).}
\label{FIGD6}
\vspace{-1mm}
\end{figure}

To better understand the low-energy physics of a system in the weak-coupling regime, we calculate the LDOS corresponding to the vertical cuts in Fig. \ref{FIGD3}(b). The results are shown in Fig. \ref{FIGD4}. There are two major features associated with weak SM-SC coupling. First, because of the weak induced pairing, there is no significant gap at zero Zeeman field and no localized states, meaning that the characteristic localization length is much larger than the size of the system. Second, since the induced disorder is negligible, there are no disorder-induced low-energy states and the spectral features are highly symmetric.

Finally, we consider the sequence of seven consecutive thickness values corresponding to ``optimal'' relative amplitudes (green dots) in Fig. \ref{FIGD1} (for $n_y=2$). For $\Lambda = 0.4~$meV, all systems are in the intermediate SM-SC coupling regime.  The dependence of the DOS on the Zeeman field for $\mu=0.5~$meV and values of $L_z$ corresponding to the sequence of interest is shown in Fig. \ref{FIGD5}. As a result of the systems being in the intermediate SM-SC coupling regime, at zero magnetic field ($\Gamma=0$) all systems have a finite induced gap of order $\sim 0.3\Delta_0-0.6\Delta_0$ (i.e., $\sim 0.1-0.2~$meV). Upon applying an external magnetic field, the lowest energy modes collapse toward zero energy at $\Gamma \sim 0.35-0.5~$meV. The behavior at higher values of $\Gamma$ corresponds to three different scenarios. Most systems, more specifically the spectra in panels (a-d) and (g), are characterized by a low-energy mode separated from the rest of the spectrum by a finite quasiparticle gap. Except panel (b), the low-energy mode exhibits relatively large energy splitting oscillations indicative of finite size effects. Indeed, the LDOS corresponding to the vertical cuts in \ref{FIGD5} (a) and (g), which is shown in \ref{FIGD6} panels (a) and (e), respectively, demonstrates that the lowest energy BdG state corresponds to a pair of modes localized near the ends of the wire and having characteristic length scales comparable to the size of the system (hence, a significant overlap). Also note that the LDOS features are highly symmetric, which implies low-effective disorder. In other words, the features characterising the spectra in \ref{FIGD5} (a-d) and (g) [also Fig. \ref{FIGD3}(c)] and the LDOS in Fig. \ref{FIGD6} (a) and (e) indicate the emergence of (precursor) topological superconductivity and Majorana bound states in a finite size (relatively short) hybrid system with weak effective disorder. Of course, increasing the length $L_x$ of such a wire will result in the emergence of well-separated, topologically-protected MZMs. 

The second scenario is illustrated by the spectrum in \ref{FIGD5}(f). Note that $L_z=9.59~$nm corresponds to the largest average Green's function amplitude in the sequence (see Fig. \ref{FIGD1}). Comparison with the results in Fig. \ref{FIG17} suggests the presence of a disorder-induced  low-energy ABS, in addition to a pair of near-zero energy Majorana modes. The corresponding LDOS, shown in Fig. \ref{FIGD6}(d) confirms this suggestion. Moreover, it shows highly asymmetric and relatively localized features characteristic of strong induced disorder. Note that the disorder-induced ABS localized near the middle of the wire has almost zero energy and is completely ``invisible'' in transport measurements. 

The third scenario is illustrated in Fig. \ref{FIGD5}(e) and corresponds to a proliferation of disorder-induced low-energy states, as confirmed by the  asymmetric features characterizing the corresponding LDOS in Fig. \ref{FIGD6} (b) and (c). We point out that the total SM-SC coupling in this system is weaker than the coupling characterising the system in Figs. \ref{FIG15} and \ref{FIG16}, which show weak disorder effects. In general, we have noticed that for the surface roughness realization investigated here, the transition from low-effective disorder to intermediate/high induced disorder occurs at lower values of $\Lambda$ (as compared to the surface roughness discussed in the main text). In other words, this surface roughness (characterized by larger in-plane and shorter transverse characteristic length scales) is less effective in enhancing the interface Green's function and suppressing its thickness dependence, but, for a given SM-SC coupling $\Lambda$, it is more susceptible to generate strong induced disorder.
This property emphasizes once again the importance of experimentally investigating the details of SC disorder and identifying ``optimal'' disorder and SC film parameters that can be reliably realized, which remain critical outstanding tasks.

\bibliography{REFERENCES}
\end{document}